\documentclass[screen, acmsmall, natbib=false]{acmart} 

\usepackage{amsmath, amssymb, amsfonts}

\usepackage{mathtools} 
\usepackage{braket}

\usepackage{algorithmic}
\usepackage{graphicx}
\usepackage{rotating}
\usepackage{adjustbox} 

\usepackage{soul}
\usepackage{xcolor}
\definecolor{lightblue}{rgb}{0.5,0.9,1}
\sethlcolor{lightblue}

\usepackage{caption}  
\usepackage{subcaption}

\usepackage[backend=bibtex, sorting=none, style=numeric-comp]{biblatex}
\setcopyright{acmlicensed}
\acmDOI{10.1145/3788677}

\newboolean{showcomments}
\setboolean{showcomments}{true}         
\ifthenelse{\boolean{showcomments}}
  {\newcommand{\nb}[2]{
  \fbox{\bfseries\sffamily\scriptsize#1}
     {\sf\small$\blacktriangleright$\textit{\textcolor{red}{#2}}$\blacktriangleleft$}
   }
  }
  {\newcommand{\nb}[2]{}
   
  }

\usepackage{hyperref}

\begin{document}

\title{\textbf{A Taxonomy of Real Faults for Hybrid Quantum-Classical Software Architectures.} \\
{\footnotesize \textsuperscript{}}
\thanks{}
}

\author{Avner Bensoussan}
\orcid{0009-0007-3285-9468}
\affiliation{
 \institution{King's College London}
  \city{London}
 \country{UK}
}
\email{avner.bensoussan@kcl.ac.uk}

\author{Gunel Jahangirova}
\orcid{0000-0002-1423-1083}
\affiliation{
 \institution{King's College London}
  \city{London}
 \country{UK}
}
\email{gunel.jahangirova@kcl.ac.uk}

\author{Mohammad Reza Mousavi}
\orcid{0000-0002-4869-6794}
\affiliation{
 \institution{King's College London}
  \city{London}
 \country{UK}
}
\email{mohammad.mousavi@kcl.ac.uk}



\begin{abstract}
With the popularity of Hybrid Quantum-Classical architectures, particularly noisy intermediate-scale quantum (NISQ) architectures in the gate-based quantum computing paradigm, comes the need for quality assurance methods tailored to their specific faults. In this study, we propose a taxonomy of faults in gate-based Hybrid Quantum-Classical architectures accompanied by a dataset of real faults in the identified categories. To achieve this, we empirically analysed open-source repositories for fixed faults. We analysed over 5000 closed issues on GitHub and pre-selected 529 of them based on rigorously defined inclusion criteria. We selected 133 faults that we labelled around symptoms and the origin of the faults. We cross-validated the classification and labels assigned to every fault between two of the authors. As a result, we introduced a taxonomy of real faults in gate-based Hybrid Quantum-Classical architectures. Subsequently, we validated the taxonomy through interviews conducted with eleven developers. The taxonomy was dynamically updated throughout the cross-validation and interview processes. The final version was validated and discussed through surveys conducted with an independent group of domain experts to ensure its relevance and to gain further insights.
\end{abstract}



\maketitle

\section{Introduction}

Quantum Computing (QC) holds the potential to enhance computational capabilities and address certain complex domain-specific problems that are intractable for classical methods~\cite{gheorghe-pop_quantum_2020}. In the last four decades, QC has evolved from theoretical concepts to real publicly available hardware and services. However, current computers are limited in their scalability to solve real-life calculations. While the main actors in realising the potential of QC have hitherto been physicists,  
computer scientists, and in particular, software engineers, are playing an increasing role in the following  
steps~\cite{miranskyy_testing_2019}. Current quantum hardware raises the  
need for specific software infrastructures and tools to mitigate its scalability issues~\cite{greiwe_effects_2023}.  

For the foreseeable future, quantum systems—specifically gate-based quantum computers—are going to be interfaced and complemented with classical computers to achieve large-scale tasks. This has led to a wide-scale adoption of a new type of software architecture called \emph{Hybrid Quantum-Classical architectures}.  
In particular, the current stage of quantum computing is termed the Noisy  
Intermediate-Scale Quantum (NISQ) era, where noisy quantum computation is wrapped around an architecture involving classical optimisation. Several algorithms have been developed  
to exploit the potential of current computers and are referred to as  
NISQ algorithms~\cite{lau_nisq_2022}, which are instances of Hybrid Quantum-Classical architectures based on the gate-model quantum computing paradigm.  
Until we have access to full-scale fault-tolerant quantum computing, we expect a quantum advantage for domain-specific NISQ  
algorithms~\cite{herrmann_quantum_2023}. However, even in the presence of fault-tolerant quantum computing, complex Hybrid-Quantum Classical architectures are likely to remain prominent because of the complementary nature of these two computing paradigms. The interest for Hybrid Quantum--Classical architectures has grown significantly over the past decade, as demonstrated by the number of related publications depicted in Figure~1. Figure~1a displays the yearly total number of publications resulting from a search query of ``Hybrid Quantum Classical'' keywords using arXiv's tool from 2010 until today, while Figure~1b shows the same trend in peer-reviewed venues (Springer, ACM). We observe a clear growth with a significant increasing trend, reflecting the rising interest and investment in hybrid architectures as a promising use case of quantum computing. This trend likely stems from a broader recognition that quantum computing will, in the near future, be embedded within existing classical infrastructures. As the field matures, hybrid software will play a crucial role in enabling the seamless integration of this new computational paradigm into current systems. Such growing interest highlights the need for tailored quality-assurance methods and tools. Indeed, Ramalho, de Souza \& Chaim note ``gaps ... related to ... input states with hybrid interfaces'' and call for ``testing and debugging techniques that take advantage of the unique quantum computing characteristics'' \cite{Ramalho2024}. Similarly, Murillo~et~al. \cite{Murillo2025} observe that specific testing methods are crucial for Quantum and Hybrid Systems to scale. . These insights underscore why hybrid-specific testing, debugging, and verification strategies are essential to understand and classify the common fault types in hybrid architectures --- a central motivation for the taxonomy presented in the following section.


\begin{figure}[hbt!]
    \centering
    \begin{subfigure}[t]{0.48\textwidth}
        \includegraphics[width=\linewidth]{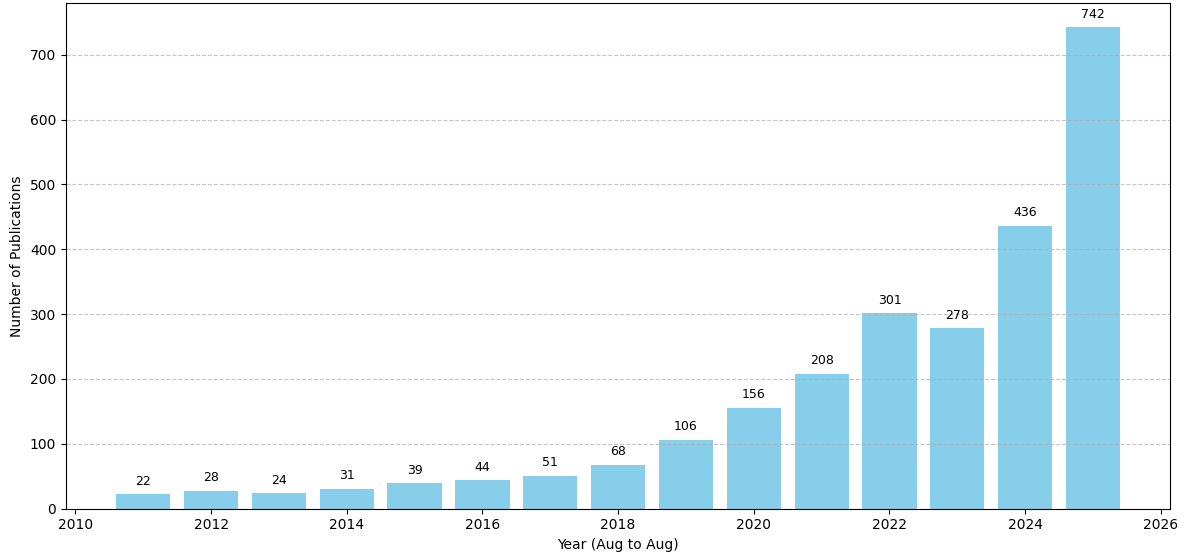}
        \caption{Publication Trend in ArXiv from 2010 to today.}
        \label{fig:trend}
    \end{subfigure}
    \hfill
    \begin{subfigure}[t]{0.48\textwidth}
        \includegraphics[width=\linewidth]{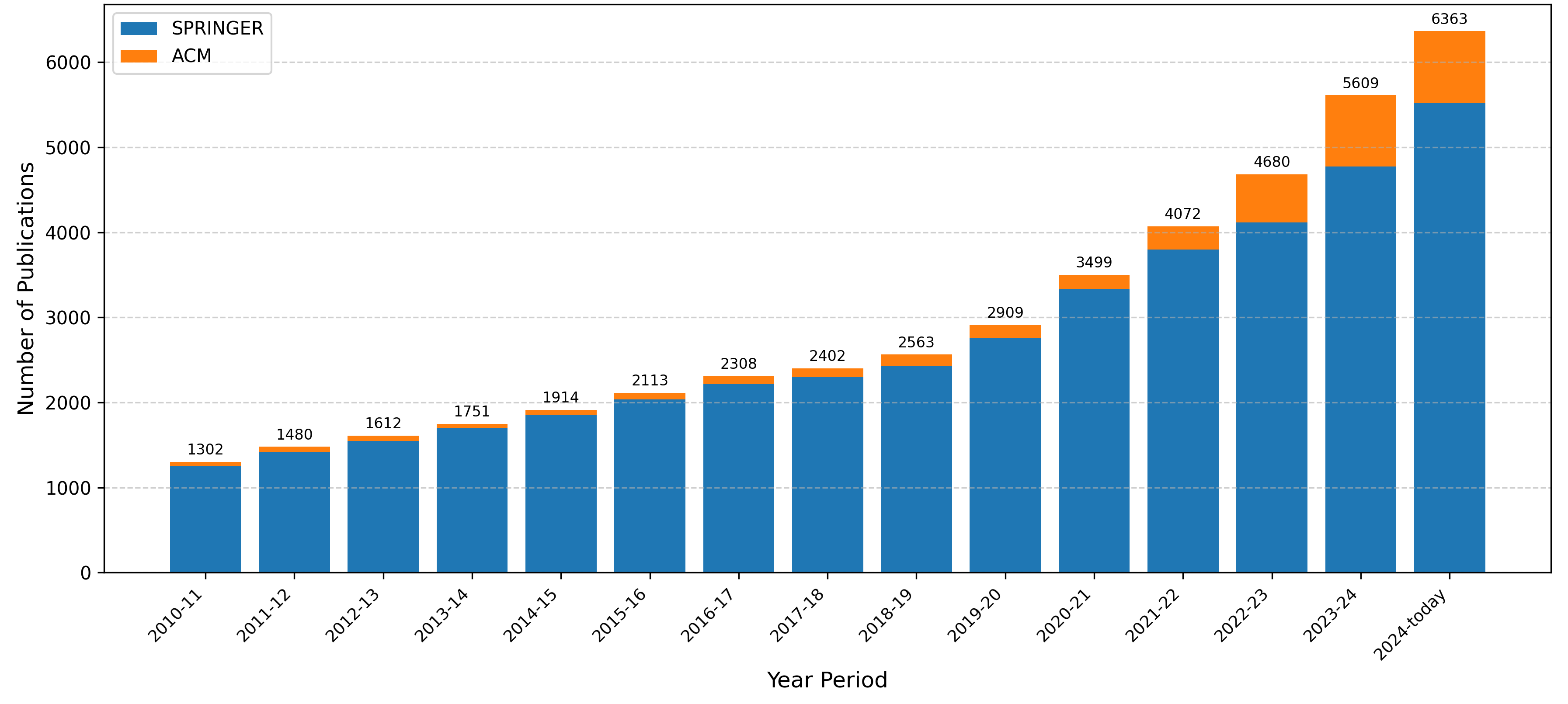}
        \caption{Publication Trend in SPRINGER (in blue) and ACM (in orange) from 2010 to today.}
        \label{fig:second}
    \end{subfigure}
    \caption{Comparaison of Publication Trends for Hybrid Quantum Classical in ArXiv and Peer-reviewed venues from 2010 to today.}
    \label{fig:two_trends}
\end{figure}

To respond to this need, in this study, we propose a first taxonomy of real faults in Hybrid Quantum-Classical architectures. Although it is focusing on NISQ algorithms running on gate-based quantum hardware, we expect our taxonomy to apply to future Hybrid Quantum-Classical architectures, potentially including fault-tolerant quantum computing. We consider NISQ algorithms, which represent an early version of Hybrid Quantum-Classical architectures, to be an interesting case study to understand the behaviour of Quantum Systems when they are embedded in Classical Computations.  
To develop this taxonomy, we considered over 5000 Github issues, preselected 529 of them using rigorously defined inclusion criteria, and analysed them between two authors. We cross-validated our results and selected 133 real faults in Hybrid Quantum-Classical architectures.  
We then structured the results into a taxonomy and further labelled the faults in each classification with more specific descriptions of the nature of the fault. The initial version of the taxonomy was validated and continuously updated through interviews with eleven domain experts. The outcome of the interviews was validated once more through surveys involving a separate group of experts.

\noindent \textbf{Research questions.}
 To our knowledge, no in-depth analysis of Hybrid Quantum-Classical architecture faults has been conducted. Hence, quantum testing research would benefit from a structured taxonomy of real faults. The following two research questions drive our study: 

\begin{itemize}\itemsep0em
\item  \textit{\textbf{RQ1:} What are typical failure causes in Hybrid Quantum-Classical architectures?} 

The integration of quantum circuits into a loop of classical calculations forms the structure of Hybrid Quantum-Classical architectures. It is natural to expect unique fault categories associated with it, pertaining to the quantum or classical components and their interfaces. We see similarities between such architectures and machine learning architectures, which motivated us to consider similar efforts in that domain~\cite{humbatova_taxonomy_2019} and analyse the similarities in the outcome of our research.  

\item  \textit{\textbf{RQ2: } How can we develop a taxonomy of real faults that captures the specific structure of Hybrid Quantum-Classical architecture?} 

We would like to investigate how real open-source faults (mined from repositories) can be complemented by real faults elicited through interviews with domain experts. We would like to see if the resulting taxonomy can be validated by independent experts and if it can lead to further insights. 

\end{itemize}

Those questions motivated our analysis of real Hybrid Quantum-Classical architectures faults. 
The contributions of this paper are three-folded.

\begin{itemize}\itemsep0em

\item  \textbf{We benchmarked real Hybrid Quantum-Classical faults.} We empirically analysed public repositories containing Hybrid Quantum-Classical architectures. We searched 5000 fixed issues and selected 133 real faults. 
\item  \textbf{We proposed a first taxonomy of real Hybrid Quantum-Classical architecture faults.} By analysing and categorising those real faults, we developed a taxonomy of real Hybrid Quantum-Classical architecture faults. 
\item  \textbf{We validated this taxonomy against other available datasets and through interviews with experienced developers working in this domain and a survey involving an independent group of experts.} To ensure the validity of our results, we compared them to previous works and presented them to experienced Hybrid Quantum-Classical architecture developers and  researchers. Their feedback was incorporated into our results to create the final taxonomy, validate the outcome, and provide reflective insights on the final results.
\end{itemize}

\noindent \textbf{Structure of the paper.} Section \ref{background} introduces quantum computing and key concepts to understand Hybrid Quantum-Classical architectures. In Section \ref{relatedwork}, we review the related work, both in classical and quantum software engineering. Section \ref{methodology} reviews the methodology of this study before presenting our results in Section \ref{results} and discussing them in Section \ref{discussion}. We state threads to the validity of our study in Section \ref{threats to validity} before concluding in Section \ref{conclusion} and suggest some avenues for future work.
\section{Background}
\label{background}

This section briefly introduces Quantum Computing and Hybrid Quantum-Classical architectures.

\subsection{Quantum Computing}
Quantum computing (QC) was conceptualised to simulate quantum phenomena based on the postulates of quantum mechanics~\cite{feynman_simulating_1982}. It was later found to have potential applications that could offer significant speed-up over its classical counterpart. The resulting potential speed-up is often referred to as "quantum supremacy"~\cite{arute_quantum_2019}. However, demonstrating quantum supremacy on real hardware remains a long-standing challenge, especially at a scale where it would solve real-life calculations. Most agree this stage of QC will likely last for the next few years if not decades, and refer to it as the NISQ era~\cite{preskill_quantum_2018}. Regardless of the trajectory to large-scale fault-tolerant computing, quantum computers will have a different type of strength than classical computers and are going to be complemented by classical computers in Hybrid architectures. 
 NISQ algorithms are a prominent example that Hybrid architecture combining small quantum circuits with classical computations could present some computational advantages~\cite{lau_nisq_2022}. Variational Quantum Algorithms (VQA) are the most common example of an efficient combination of a reduced quantum circuit inside a classical optimisation loop~\cite{tilly_variational_2022}. We present VQA as an example of Hybrid Quantum-Classical architecture in more detail at the end of this section.

 \paragraph{Different paradigms} In the landscape of quantum computing paradigms, gate-based quantum computing stands out as the most widely adopted and a general-purpose model. It relies on the manipulation of qubits through a sequence of quantum gates, analogous to logic gates in classical computing, and forms the foundation for most current quantum algorithms, including hybrid approaches such as Variational Quantum Algorithms (VQAs). In contrast, quantum annealing is a specialized model primarily designed for solving combinatorial optimization problems by leveraging quantum fluctuations to explore low-energy states of a system. Commercial implementations, such as those from D-Wave Systems, are based on this paradigm but are not universal and are limited in their applicability beyond specific problem classes. Another alternative is measurement-based quantum computing (MBQC), which executes computations by performing a series of measurements on a highly entangled initial resource state (typically a cluster state), rather than applying gates directly. While MBQC provides a theoretically equivalent model to gate-based systems, it is less mature in terms of hardware support and algorithmic development. Given the broader applicability and better compatibility with near-term hardware, gate-based quantum computing remains the dominant framework, particularly for NISQ-era hybrid algorithms.

\subsection{Quantum Circuits}
The core of a Hybrid Quantum-Classical architecture is a Quantum circuit. A quantum circuit comprises quantum gates operating on qubits, the quantum analogue to classical bits.  Two distinctive features of qubits are that they may feature a superposition of states,  represented by a linear combination of states, and entanglement, where several qubits form a composite system such that measurement on one determines the state of the other. 

\paragraph{Composition of a circuit}
A quantum circuit used in a Hybrid Quantum-Classical architecture typically begins with the initialisation of qubits and ends with their measurement, where the information is translated into classical bits. The purpose of a quantum circuit is to harness the principles of quantum mechanics, such as superposition and entanglement, to perform computational tasks that would be infeasible or significantly slower for ``quantum parallelism'', i.e., ``evaluating a function on many different values simultaneously''~\cite{nielsenChuang10}.

\paragraph{Quantum Gates} Quantum circuits are constructed using quantum gates, which are unitary operations acting on one or more qubits. These gates manipulate the quantum state by exploiting quantum phenomena. Common single-qubit gates include the Hadamard (H) gate, which creates superposition, the Pauli-X gate, which acts like a classical NOT, and phase-shift gates such as S and T. Multi-qubit gates, such as the Controlled-NOT (CNOT) gate, are essential for creating entanglement between qubits. Together, these gates form a universal set capable of implementing any quantum computation when combined in appropriate sequences.

\begin{figure}[hbt!]
\includegraphics[width=0.5\linewidth]{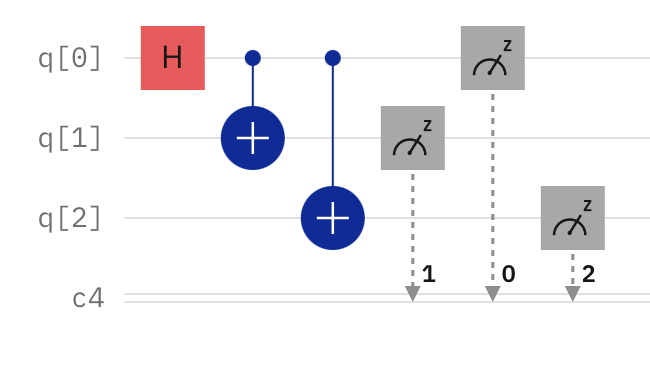}
\caption{Simple Quantum circuit creating a GHZ state.}
\label{ghz}
\end{figure}

\paragraph{Example of a quantum circuit} Fig.~\ref{ghz} represents a simple quantum circuit, creating a three-qubit Greenberger-Horne-Zeilinger (GHZ) state, which is a specific entangled quantum state of three qubits. The GHZ state signifies that the three qubits are entangled in such a way that if the state of one qubit is measured, the states of the other two qubits are also determined. Input variables $q[0]$, $q[1]$, and $q[2]$ represent the 3 input qubits. The horizontal line represents time, the square and round elements are quantum logic gates, and the final boxes and arrow elements represent measurements, where the qubit is observed and its state is extracted into classical bits. The maximum number of gates along all horizontal lines in a circuit is called depth. Fig.~\ref{ghz} represents a circuit of depth 3. \\

\paragraph{Quantum Noise}
There are different sources of noise in quantum computing, including unwanted entanglement among particles and imperfection in gates. 
Noise accumulates throughout the quantum circuit~\cite{clerk_introduction_2010}. Any quantum system suffers decoherence, induced by the interactions with its environment, causing it to lose its quantum behaviour over time~\cite{lau_nisq_2022}. Hence, current Hybrid Quantum-Classical architectures, e.g., NISQ algorithms, aim to reduce the number of qubits and the gate depth of the circuit to keep the noise level manageable. \\

\subsection{Hybrid Quantum-Classical Architectures}
In this section, we take a closer look at NISQ algorithms as prominent examples of Hybrid Quantum-Classical architectures currently in use, particularly, in open-source repositories for quantum computing. 






\paragraph{Structure of Hybrid Quantum-Classical Architectures} The general workflow consists of four main phases:

\begin{enumerate}\itemsep0.6em

\item \textit{Classical pre-processing.} 
The problem of interest (e.g., optimisation, simulation, or classification) is mapped into a quantum-compatible representation. This typically involves formulating a cost function and encoding it into a parameterised quantum circuit, known as an Ansatz. Initial parameters are chosen—often heuristically or randomly—and qubits are prepared in a standard reference state such as \( \left|0\right\rangle^{\otimes n} \).

\item \textit{Quantum circuit execution.} 
A parameterised quantum circuit is executed with the current parameter values. The circuit consists of layers of single-qubit rotations and entangling gates and is typically kept shallow to mitigate noise. This is the core of a Hybrid Architecture. It typically has a low gate depth and few qubits to manage the noise level. Current Hybrid architectures utilise around 50 qubits, and 1000 gates at maximum, with an average gate depth of 20~\cite{lau_nisq_2022}. The circuit is run multiple times (shots) to generate measurement statistics.

\item \textit{Classical optimisation loop.}
Measurement outcomes are used to estimate the value of the cost function, which is then fed into a classical optimiser that updates the parameters. This process iterates, forming a hybrid feedback loop until a convergence criterion is met.

\item \textit{Classical post-processing.}
Once a sufficiently good solution is found, the output—such as optimised parameters or low-energy states—is further processed or interpreted using classical resources, depending on the application.

\end{enumerate}

\paragraph{Variational Quantum Algorithms (VQAs)}
\label{vqa seq}
VQAs are among the most promising examples of NISQ algorithms \cite{tilly_variational_2022}. 
The main goal of a VQA is to find the optimal parameters for a parameterized quantum circuit, leading to a solution for a given computational problem. We provide an overview of the structure and functioning of a VQA. This structure is typical of Hybrid Architecture. Future Fault-Tolerant Quantum Computing could replace current noisy hardware.

\begin{enumerate}\itemsep0em

\item[1.]  \textit{Problem Definition.} The first step involves defining a computational problem that can benefit from quantum computing and translating it into an objective or cost function. In most applications, it consists of a Hamiltonian representation (a matrix representing system state evolution), connecting the problem domain to the energy level of a quantum system.

\item[2.]  \textit{Optimisation Process.} A parameterized quantum circuit, known as the variational Ansatz, is designed. This circuit contains gates with adjustable parameters, denoted as $\theta$. It is initialised with an input $\theta_0$. A quantum state $\ket{\psi(\theta)}$ is measured, and the outcomes are used to compute the expectation value of the objective function $O(\theta, \{\braket{H}_{U(\theta)}\})$.

\item[3--4.]  \textit{Convergence Check and Output.} The optimization process continues until a convergence criterion is met, indicating that further iterations are unlikely to significantly improve the solution. This convergence check ensures that the algorithm has reached a stable and potentially optimal solution. The final set of optimized parameters $\theta_{opt}$ represents the solution to the quantum problem. This solution can be used for further analysis or as the output of the VQA.

\end{enumerate}

\begin{figure}[hbt!]
\includegraphics[width=0.8\linewidth, scale=0.2]{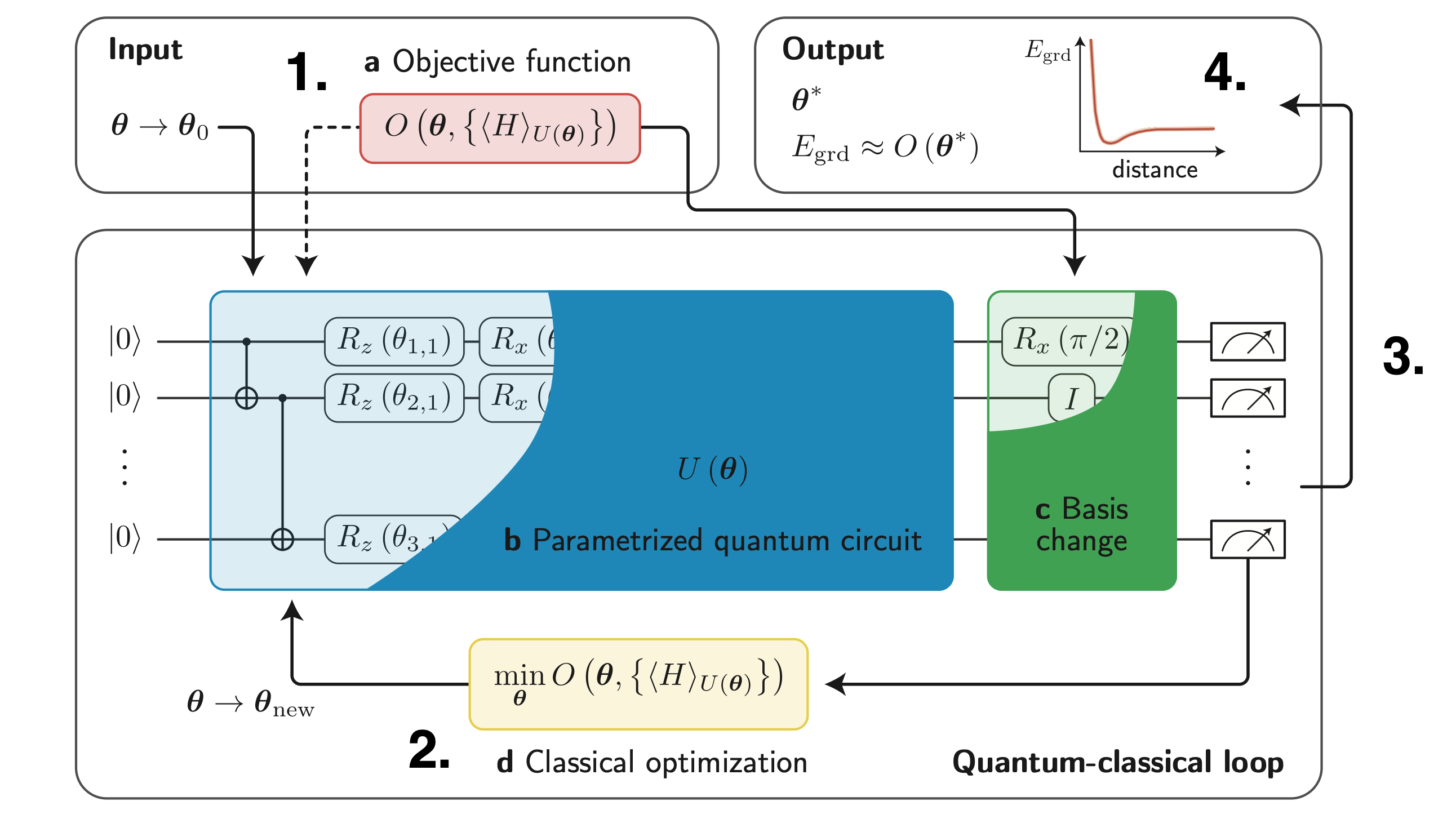}
\caption{Detailed structure of a VQA \cite{bharti_noisy_2022}}
\label{VQE}
\end{figure}

An in-depth explanation and a detailing figure can be found in the work of Bharti et al. \cite{bharti_noisy_2022}. Fig.~\ref{VQE} \cite{tilly_variational_2022} is a diagrammatic representation of a VQA in which each of the described steps can be found.

\section{Related work}
\label{relatedwork}

This section positions our study within the broader context of software engineering and quantum computing research. We begin by reviewing established practices in classical software engineering, including the construction of real fault benchmarks and the development of taxonomies for fault classification. We then examine existing efforts in quantum software testing, with a particular focus on real faults in quantum and Hybrid Quantum-Classical architectures. Finally, we compare our study to the most closely related work in this area. A summary of these contributions and their limitations is presented in Table~\ref{tab:relatedwork}, highlighting the novelty and gap addressed by our work.

\subsection{Classical Computing}
\paragraph{Bug Benchmarks} 
Building datasets of reproducible real faults is a common and valuable practice in classical software engineering. These datasets offer a more realistic and eclectic base for software testing research on artificially generated faults. Noticeably, Defects4J gathers reproducible real faults in Java \cite{just_defects4j_2014}. It was used in several studies, for instance, to inform the design of testing frameworks \cite{martinez_automatic_2017} and to experiment with software engineering techniques \cite{aleti_defects4j_2020}. We expect our study could similarly benefit research in Hybrid Quantum-Classical architectures and serve as an experimentation base for future studies. \\

\paragraph{Use of Taxonomy in Software Engineering}
In science and engineering, systematically organising and classifying knowledge advances the field by providing a common discourse; it facilitates knowledge sharing, clarifies relationships between concepts, identifies knowledge gaps, and supports decision-making. This classification helps researchers and practitioners generalise, communicate, and apply findings effectively. As stated in Usman et al.'s systematic mapping study of Taxonomies in Software Engineering (SE) \cite{usman2017taxonomy}, SE being a broad and diverse field, it relies heavily on taxonomies structured classification schemes that describe terms and their relationships to organise its extensive knowledge base. In summary, taxonomies contribute to both theoretical clarity and practical applicability in the software engineering discipline.

\paragraph{Real Faults Taxonomy}
Taxonomies being a common way to propose a structured overview of a given problem in software engineering \cite{UsmanEtAl017}, a study that inspired our work proposes a taxonomy of real faults in deep learning systems \cite{humbatova_taxonomy_2019}. They analysed GitHub commits and issues for popular DL frameworks and related Stack Overflow posts to propose a taxonomy of real DL faults. They validated their work by interviewing 20 developers and integrating feedback received to their results. Our work follows a similar methodology applied to Hybrid Quantum-Classical architecture. Building a fault dataset and a taxonomy is a process that can be transferred to a different branch of software engineering since the data available online is similar. Also, the developers' interview structure can be easily adapted to another specialty.

\subsection{Quantum Computing}
\paragraph{Real Faults in Quantum and Hybrid architectures}
Quantum software testing is challenging. Several studies point to a growing need for developing testing and debugging tools specific to quantum programs \cite{miranskyy_testing_2019}\cite{paltenghi_bugs_2022}\cite{metwalli_tool_2022}\cite{gill_quantum_2022}\cite{pontolillo_multi-lingual_2023}. Some benchmarks of quantum bugs and bug fixes are available in the literature \cite{zhao_bugs4q_2021}\cite{campos_qbugs_2021}\cite{luo_comprehensive_2022}, as well as a low-level Quantum Benchmark Suite, QASMBench, focusing on
NISQ Evaluation and Simulation \cite{li_qasmbench_2023}. Zhao et al. \cite{zhao_identifying_2021} identified some bug patterns in Quantum Programs after an in-depth analysis of Qiskit programs. In our study, we used one of them, namely, the Bugs4Q dataset \cite{zhao_bugs4q_2021} to validate our search process. Although Bugs4Q has a broad scope and does not focus on Hybrid Quantum-Classical architectures, it contains one NISQ bug. In their systematic literature review, Gill et al. \cite{gill_quantum_2022} briefly mention Hybrid Quantum-Classical architectures as a priority research area, without mentioning testing techniques for them. To our knowledge, real faults in Hybrid Quantum-Classical architectures, such as NISQ algorithms,  have not been investigated yet. The closest study to ours concerns faults in Quantum Machine Learning, which is a special case of a NISQ algorithm \cite{zhao_empirical_2023}. We used both datasets to validate our search query and process. We notice a growing interest in recent literature in Quantum testing and quality assurance, and very recently some studies have started to emphasise their investigation around NISQ and Hybrid Quantum-Classical architectures. 

\paragraph{Other taxonomy}
Parallel to our work, Zappin et al. \cite{zappin_when_2024} recently released a comprehensive study characterizing
Hybrid Quantum-Classical issues discussed in
Developer Forums, which represents the closest work to ours to date. However, they solely investigated discussions available on Xanadu Discussion Forums and on QC Stack Exchange. The issues they used do not originate from GitHub and do not follow the strict inclusion criteria we used as described in Section \ref{methodology}, hence there is no overlap between our two works. Their study gave an overview of current causes of issues in a software engineering perspective, gathering them around the following categories: Software Faults, Hardware/Simulator issues, Configuration issues, Developper Errors, Library and Plateform Issues. Our study characterises faults in Hybrid Quantum-Classical architectures around their structuring components, and identify their typical weaknesses, aiming to further understand their nature and stay relevant to future versions of such architectures. These two taxonomies are therefore complementary. They were conducted in parallel, without consulting one another, which highlights once more the relevance of such a work.

\subsection{Comparative Analysis of Our Study}

\paragraph{Data Source}
A key strength of our study lies in the diversity and rigor of its data sources. While most previous datasets either focus exclusively on synthetic benchmarks such as QASMBench~\cite{luo_comprehensive_2022,li_qasmbench_2023}, on general-purpose quantum software like Bugs4Q~\cite{zhao_bugs4q_2021} and QBugs~\cite{campos_qbugs_2021}, or on community-reported developer issues without validation such as the recent work by Zappin et al.~\cite{zappin_when_2024}, our study draws on real-world Hybrid Quantum–Classical architectures faults mined directly from GitHub repositories verified and fixed by developers. We complemented this empirical evidence with curated expert knowledge obtained through targeted interviews and surveys. This multi-pronged data collection strategy allows our taxonomy to be grounded in both reproducible faults and practical insights from developers and researchers actively engaged in Hybrid Quantum-Classical architectures design and deployment.

\paragraph{Focus}
Prior studies have focused on specific domains such as classical software engineering (e.g., Defects4J~\cite{just_defects4j_2014}), deep learning systems (e.g., Humbatova et al.~\cite{humbatova_taxonomy_2019}; Thomas et al.~\cite{thomas2024muprl}), or specialised subfields such as Quantum Machine Learning (e.g., Zhao et al.~\cite{zhao_empirical_2023}). In contrast, our work targets the emerging domain of Hybrid Quantum–Classical architectures, where quantum and classical components interact within the same computational workflow. This focus reveals new categories of faults that cannot be adequately captured by existing taxonomies developed for purely classical or purely quantum contexts. By centring on real Hybrid architectures, our taxonomy captures the distinctive fault patterns and interaction failures characteristic of hybrid systems, providing researchers and practitioners with a timely and practically relevant foundation for advancing testing and debugging methodologies in Hybrid architectures engineering.

\paragraph{Validation}
A defining strength of our work lies in its rigorous validation methodology. Whereas prior studies have typically relied on manual curation (e.g., Bugs4Q~\cite{zhao_bugs4q_2021}; QBugs~\cite{campos_qbugs_2021}) or literature-based triangulation (e.g., Gill et al.~\cite{gill_quantum_2022}), our approach introduces a more systematic, multi-stage validation pipeline. This pipeline combines cross-validation with existing datasets, triangulation across heterogeneous sources—including GitHub issue reports, developer interviews, and targeted surveys—and expert assessment obtained through both interviews and a dedicated round of validation surveys. This methodological depth ensures that our taxonomy is empirically grounded and verified for accuracy and practical relevance by a diverse set of practitioners and researchers across multiple quantum-computing subfields. To reduce false positives (i.e., issues that do not correspond to actual faults), we restricted our dataset to closed issues, guaranteeing that each entry reflects a confirmed and resolved fault. This choice enhances the reliability and reproducibility of our results and aligns with best practices adopted in previous peer-reviewed software-fault datasets (Zhao et al., 2021~\cite{zhao_bugs4q_2021}; Campos et al., 2021~\cite{campos_qbugs_2021}; Luo et al., 2022~\cite{luo_comprehensive_2022}; Li et al., 2023~\cite{li_qasmbench_2023}; Zappin et al., 2024~\cite{zappin_when_2024}).

\paragraph{Limitation}
Despite these strengths, we acknowledge that our taxonomy, while broad and empirically grounded, is inherently shaped by the current state of hybrid quantum-classical technologies and the faults that have been reported and documented to date. As quantum hardware, software frameworks, and hybrid orchestration techniques evolve, new classes of faults will inevitably emerge. This limitation underscores the importance of designing our taxonomy to be extensible and adaptable to future architectures — a design choice that we have explicitly prioritised to ensure its long-term relevance and usefulness to the research and development community.

\begin{table}[ht]
\centering
\caption{Summary of Related Work and Positioning of Our Study}
\label{tab:relatedwork}
\small
\begin{adjustbox}{max width=\textwidth}
\begin{tabular}{p{3.8cm}|p{3.0cm}|p{3.4cm}|p{4.4cm}|p{5.2cm}}
\hline
\textbf{Study / Dataset} & \textbf{Data Source} & \textbf{Focus} & \textbf{Validation Steps} & \textbf{Gap / Limitation} \\ \hline

Defects4J~\cite{just_defects4j_2014} & Java GitHub Repositories & Real faults dataset for testing and debugging & Extensive dataset curation and cross-validation with multiple Java projects & Not applicable to quantum or Hybrid architectures \\ \hline

Humbatova et al.~\cite{humbatova_taxonomy_2019}, Thomas et al.~\cite{thomas2024muprl} & GitHub, Stack Overflow & Taxonomy of real faults in deep learning applications & Expert-based evaluation of taxonomy; empirical validation on DL repositories & Does not cover quantum or Hybrid software architectures \\ \hline

Zhao et al. (Bugs4Q)~\cite{zhao_bugs4q_2021} & GitHub (Qiskit) & Benchmark of bugs in quantum programs & Manual curation of issues and patches from Qiskit repositories & Contains only a single hybrid-related bug; not tailored to Hybrid architectures \\ \hline

Campos et al. (QBugs)~\cite{campos_qbugs_2021} & GitHub (Qiskit, etc.) & Catalog of quantum software bugs & Validation through comparison with prior datasets and literature review & Focuses on general quantum software; lacks hybrid specificity \\ \hline

Zappin et al.~\cite{zappin_when_2024} & Developer Forums (e.g., Xanadu, StackExchange) & Developer-reported issues and discussions & Descriptive qualitative validation via discussion analysis; no systematic cross-validation & Lacks GitHub-based empirical faults; no validation or fault taxonomy \\ \hline

Gill et al.~\cite{gill_quantum_2022} & Literature Review & Research priorities in quantum software engineering & Literature triangulation and expert interpretation & Mentions Hybrid architectures, but lacks empirical fault analysis and testing focus \\ \hline

Zhao et al.~\cite{zhao_empirical_2023} & GitHub (QML Frameworks) & Empirical analysis of quantum machine learning software faults & Manual inspection of commits and issue reports; triangulation with literature & Concentrated on QML; not representative of general hybrid algorithms \\ \hline

Miranskyy et al., Paltenghi et al., Metwalli et al., Pontolillo et al.~\cite{miranskyy_testing_2019,paltenghi_bugs_2022,metwalli_tool_2022,pontolillo_multi-lingual_2023} & Various (Literature, Tools) & Highlight challenges in quantum software testing & Conceptual validation through expert discussions and case studies in prior literature & Do not address real-world faults or hybrid-specific taxonomies \\ \hline

Luo et al., Li et al. (QASMBench)~\cite{luo_comprehensive_2022,li_qasmbench_2023} & GitHub, Benchmarks & Dataset of quantum bugs and low-level NISQ benchmark suite & Benchmark validation against standard quantum workloads and reproducibility studies & Focus on bug benchmarking and circuit-level simulation; not hybrid-focused \\ \hline

\textbf{Our Study} & \textbf{GitHub + Manual Curation + Interviews and Survey} & \textbf{Taxonomy of real faults in Hybrid architectures} & \textbf{Cross-validation, validation with other datasets, expert interviews, and expert survey-based validation} & \textbf{First empirical taxonomy based on real hybrid faults; grounded in current implementations and extensible to future architectures} \\ \hline

\end{tabular}
\end{adjustbox}
\end{table}

\paragraph{Conclusion}
In summary, compared to prior datasets and taxonomies (as outlined in Table~\ref{tab:relatedwork}), our study distinguishes itself through its diverse and reliable data sources, its focused attention on hybrid architectures, and its rigorous validation process that integrates expert perspectives with empirical evidence. By addressing the current gap in hybrid fault taxonomies and providing a reproducible dataset enriched by expert insight, our work lays a foundational basis for advancing testing strategies and improving the robustness of Hybrid quantum-classical software systems. We further detail the practical applications and future directions of our taxonomy in Section~\ref{sec:usage-guidelines}.

\section{Methodology}
\label{methodology}

\begin{figure}[htb!]
  \includegraphics[width=\linewidth, scale=0.5]{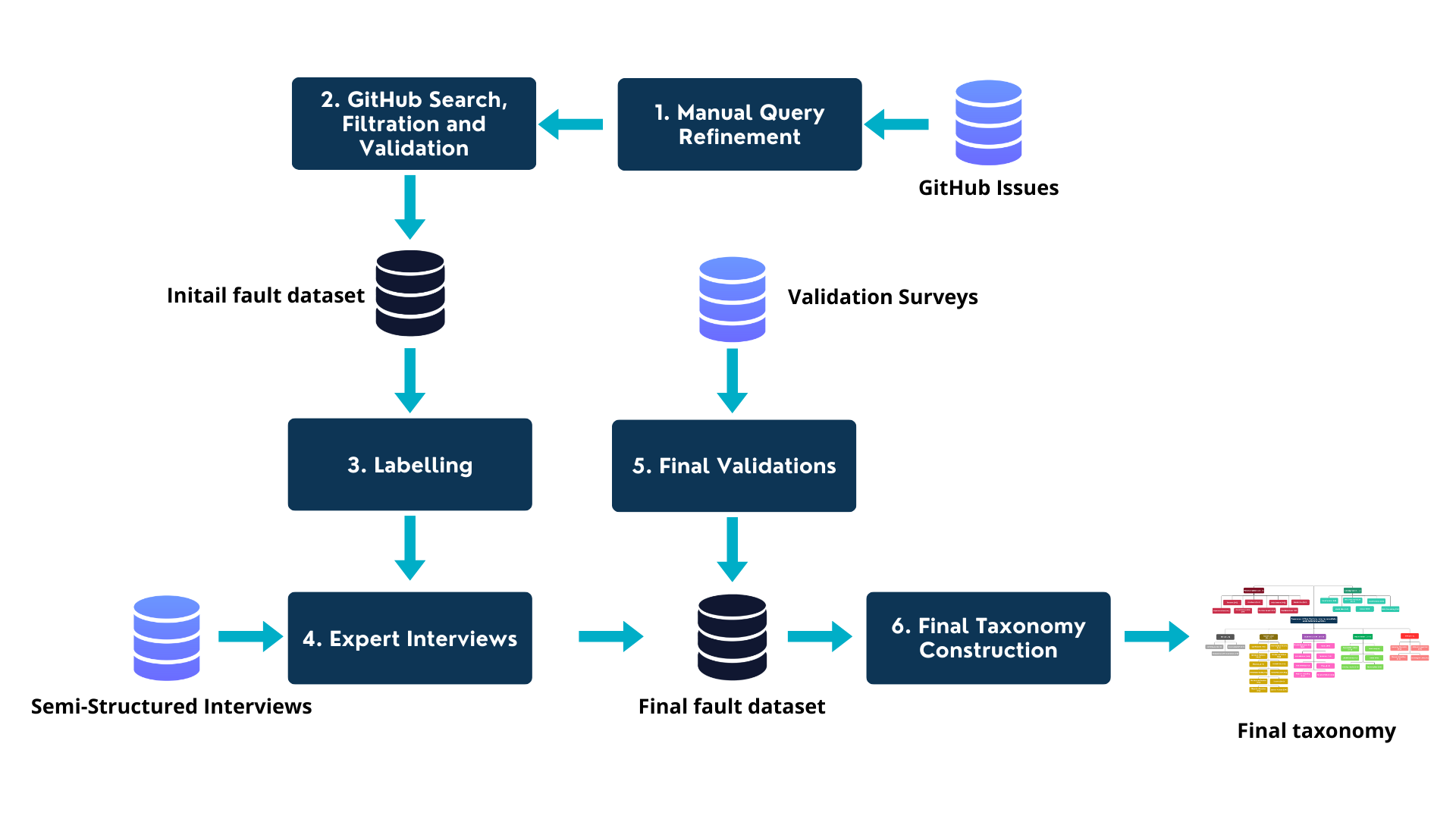}
  \caption{Overview of the 6 steps that led to the construction of our Final Taxonomy.}
  \label{fig:methodology}
\end{figure}

Fig.\ \ref{fig:methodology} presents an overview of our methodology to construct the final taxonomy. To ensure the scientific rigour of our process, we incorporated several validation steps. These included a comprehensive cross-validation of our fault dataset by the two independent authors; expert interviews to diversify the origins of our dataset, with transcripts analysed separately and then discussed jointly by the authors; and a separate survey with an independent group of experts, used solely as a final validation layer. The two authors involved in the cross-validation are a P.h.D. student specialising in Quantum Computing and a Lecturer with expertise in Machine Learning and Software Engineering. They came from different institutions and have distinct backgrounds, ensuring their independence. Each validation step was carefully documented and is fully accessible in the \href{https://doi.org/10.5281/zenodo.18172880}{replication package}. We detail the methodology in the remainder of this section based on the steps depicted in Fig.\ \ref{fig:methodology}.

\subsection*{Step 1. Manual Query Refinement}

In this step, we define the inclusion/exclusion criteria (1.1), identify the sources of real and open code bases with available bugs (1.2), design our search query (1.3), and analyse the initial search results to refine the search query (1.4).

\paragraph*{1.1. Inclusion Criteria} 
We define our inclusion criteria, based on our research questions, as follows. A fault is included in our dataset if it:

\begin{enumerate}

\item Occurs in an implementation of a Hybrid Quantum-Classical architectures (e.g.,  NISQ algorithms) (RQ1).
\item Includes a suitable description of the fault and the code associated with it (RQ1 and RQ2).
\item Includes a suitable description of the fix and the code associated with it (RQ1).
\item Includes a description of the problem and its solution in English (RQ1 and RQ2). 

\end{enumerate}

We aim to comply with ACM guidelines for reproducibility \cite{noauthor_artifact_nodate}, which means a different team should be able to reproduce these faults using the same setup.\\

\paragraph*{1.2. Analysing Available Resources} 
Quantum computing is a young field. Access to quantum simulators and real computing platforms has been available for only a few years.  Hence, the amount of open software and associated data available online is limited. We first analysed the typical web resources for such software and data: we searched Github, StackOverflow, StackExchange, and quantum-specific forums such as Pennylane and TensorFlow Quantum forums. We found results of interest for this study only on GitHub. At the time of the study, StackOverflow and StackExchange had very few discussions regarding Hybrid Quantum-Classical architectures. Those discussions were exclusively theoretical, with no code provided, or questions related to installing or importing software. Due to the inclusion criteria number 2 and 3, we excluded them. Some discussions reported faults in quantum circuits that we excluded following criteria 1 since we found only quantum faults and not Hybrid Quantum-Classical architecture faults. GitHub was, hence, the only available public source featuring source code and data satisfying our inclusion criteria. We also decided to focus on the main repositories containing Hybrid Quantum-Classical issues. This includes four famous quantum simulation platforms - Qiskit, Tensorflow Quantum, PennyLaneAI, NVIDIA Cuda-Quantum, and four independent established repositories from universities and companies - qiboteam, AgnostiqHQ, cuda-tum, goodchemistryco. This variety of fault sources aims to provide an eclectic overview of Hybrid Quantum-Classical architecture faults to answer our research questions. \\

\paragraph*{1.3. Designing a Search Query} 
We used GitHub search API to find Hybrid Quantum-Classical architecture faults and manually assess them. After an initial manual search investigation, we discovered that NISQ algorithms are the only Hybrid Quantum-Classical architecture for which there are examples of public faults and bug fixes satisfying our criteria. We first searched 'NISQ' and 'Noisy intermediate-Scale Quantum'. We found these keywords to be too generic. They mostly led to discussion around research papers, or theoretical questions about the nature of NISQ, with a vast majority of results not written in English. All the 718 results were directly excluded by at least one of our criteria, and often all four of them. Hence, we decided to search for specific algorithms designed for Hybrid Quantum-Classical architectures. 

Most implementations require importing a package named after the algorithms. For example, in qiskit, we import a class named \textit{VQE} to implement VQEs. Any source code including such an import, or any discussion mentioning the searched algorithms will be identified by this method. We, therefore, used the following keywords, which are the available algorithms according to a recent survey \cite{bharti_noisy_2022}: \textit{VQE}, \textit{VQA}, \textit{Variational Quantum eigensolvers}, \textit{Variational Quantum algorithm}, \textit{Quantum Annealing}, \textit{Gaussian boson sampling}, \textit{Analog quantum simulation}, \textit{Digital-analog quantum simulation and computation}, \textit{Iterative quantum assisted eigensolver}, \textit{Quantum Approximative Optimisation Algorithms}, \textit{QAOA}, \textit{Quantum Machine Learning}, \textit{QML}, \textit{Tensorflow Quantum}, and \textit{TFQ}. 


\subsection*{Step 2.  GitHub Search, Filtration, and Validation}
In this step, we performed the search and stored the results (2.1);  applied our inclusion/exclusion criteria and performed first the manual filtration of the results by one author and then their labelling by two authors. 
Table \ref{tab:searchoverview} provides an overview of the number of issues from our initial search process (in the column marked Initial)  and after each stage of cross-validation (columns marked "Filtration" and "Labelling"). 


\paragraph*{2.1. Organising the Search Process} 
To ensure the reproducibility of our results, we implemented a script that automatically extracts all results of a GitHub search, with the issue's title and link, into an Excel spreadsheet. The script is flexible and can easily be adapted to other queries. The results of each query are extracted into a separate sheet. Since GitHub search can only display up to 1000 results, our script automatically divides a search into sub-searches if more than 1000 results are found, and gathers all of them. As shown in Table \ref{tab:searchoverview}, the script extracted 5072 issues.


\paragraph*{2.2. Manual Filtration} We then excluded any issue that does not comply with one or more of our criteria: it does not concern a Hybrid Quantum-Classical architecture, it does not provide code or fix, or it is not in the English language. The manual filtration process was carried out by the first author, who opened and examined each of the 5072 issues extracted by the script individually. For each issue, the author read the title, body, and available code snippets (if any), and excluded any issue that clearly violated one or more of the inclusion criteria—such as being unrelated to Hybrid Quantum-Classical programming, lacking technical content, or being written in a non-English language. This step aimed to discard only obviously irrelevant issues, so as to preserve sensitivity. In cases where there was any ambiguity or doubt regarding the relevance of an issue, it was retained for further discussion and validation during the cross-review process between both authors in the next stage. As a result of applying the inclusion criteria, 529 faults remained in the dataset, as reported in column "Filtered" in Table 2.

We note that this filtration followed a conservative benchmarking principle: exclude only those issues that were clearly non-qualifying, and defer ambiguous cases to the second-stage validation. The results of manual filtration show that ‘QML’ resulted in a high number of false positives. The reason behind this is that ‘QML’ stands not only for ’Quantum Machine Learning’, but also for ’QT Modeling Language’, which typically uses a file type ‘.qml’. Out of the obtained 2861 results, only 62 were related to quantum programming, and only 6 were about Hybrid Quantum-Classical architectures. Pennylane, which is a common framework for quantum machine learning, is often imported using ‘import pennylane as qml’ as advised in their official documentation. Any program using this import without implementing a Hybrid Quantum-Classical architecture is caught by searching ‘QML’. Also, all remaining 6 issues were caught by other keywords. 

Our results also indicate that the abbreviations such as VQA instead of Variational Quantum Algorithms lead to fewer false positives, as using the full names of algorithms typically leads to issues the majority of which lack any source code.

\begin{table}[hbt!]
    \centering      
    \caption{Number of identified results after each stage: ``Initial'' refers to the initial number of results obtained through GitHub search, ``Filtered'' refers to the number of issues that satisfy the inclusion criteria, ``Labelled'' refers to the number of issues that were decided to be relevant during the manual labelling process by two authors.}
 
    \begin{tabular}{c|c|c|c}
        \hline
        \textbf{Query} & \textbf{Initial} & \textbf{Filtered} & \textbf{Labelled} \\
        \hline
        VQE & 792 & 162 & 28  \\
        \hline
        VQA & 1886 & 0  & 0  \\
        \hline
        QAOA & 466 & 87  & 27 \\
        \hline
        QML & 2861  & 6  & 0   \\
        \hline
        TFQ & 226 & 36 & 19  \\
        \hline
        Quantum Machine Learning  & 1038 & 164 & 20 \\
        \hline
        Variational Quantum algorithm & 100 & 10 & 1  \\
        \hline
        TensorFlow Quantum&  433 & 66 & 22  \\
        \hline
        Other Queries & 131 & 4 & 16  \\
        \hline
        \textbf{TOTAL}& 5072 & 529 & 133  \\
        
    \end{tabular}
    \label{tab:searchoverview}
\end{table}

 
        

\subsection*{Step 3: Labelling}
In this step, we first perform a pilot review of the faults to define a structure for coding more information about them (3.1); we use this structure to code information about each fault, including their type, symptoms, and potential causes   (3.2); then we perform a cross-validation of the labels (3.3). 

\paragraph*{Independence of the two authors} To mitigate potential bias arising from shared research affiliation, we implemented several measures to ensure independence and objectivity in the fault classification process. The two authors involved came from distinct disciplinary backgrounds (software engineering and quantum computing), had no prior working relationship at the time of the study, and conducted independent reviews and labelling of fault instances before resolving discrepancies through structured discussion. The classification task was inherently technical and systematic, with predefined labels and coding schemes, minimizing subjective interpretation. Furthermore, the resulting taxonomy underwent external validation through expert interviews and a follow-up survey, providing additional layers of triangulation and verification.

\paragraph*{3.1. Pilot Fault Review}
Each included fault was analysed by two different authors to identify its type, i.e., its place in the Hybrid Quantum-Classical architecture, its symptoms, and its possible root causes (indicated by the fix). We labelled the faults between two different authors to reduce the risk of manual error and individual bias.

For this pilot, we selected 30 bugs. Once each author completed the labelling process, we conducted a consensus meeting. During this pilot study, 20 conflicts emerged and were resolved through discussion.

\paragraph*{3.2. Individual Labelling}


During the discussion, we also defined high-level categories along which we decided to classify the bugs we analysed:

\begin{itemize}\itemsep0em

\item  \textbf{Parametrisation.} Fault related to the parameters of the circuit or the tensors.  
\item  \textbf{Ansatz.} Issues arising during the design of the Ansatz. Initially, this category was thought of as a sub-category of Quantum Circuits, but we decided to gather all faults specific to the Ansatz design since it is a crucial part of Hybrid architectures.
\item  \textbf{Quantum Circuit.} Faults occurring within the quantum circuit. This gathers all quantum faults not directly linked to the Ansatz.
\item  \textbf{Optimisation.} Faults encountered during the optimization loop. 
\item  \textbf{Measurement.}  Faults occurring during a measurement or a sequence of measurements.
\item  \textbf{GPU integration.}  Issues related to GPU integration.
\end{itemize}

These categories correspond to the main components of the workflow of a Hybrid Quantum-Classical architecture development. We used them only for guidance and added new categories and their subcategories when necessary. A more detailed description of the final list of these categories and subcategories is provided in Section \ref{vqa seq}.

For each bug, we extracted information on whether it is \textit{relevant} to the scope of our study or not, assuring they follow our inclusion criteria. Particularly, we evaluate if the fault was indeed reflecting an aspect of a Hybrid architecture. If it is relevant, we then identified whether this bug takes place \textit{inside a library} designed to be used when building an architecture (for example, a bug inside Qiskit libraries), or whether the bug happens due to a \textit{developer error} when using the library. We differentiate between these scenarios because when the issue lies within the library, it suggests a need for fixing or improving the library code. In contrast, if the issue is due to a developer error, then the fix should take place in the developer code. Getting an insight on which scenario takes place more often, can guide the future efforts required to prevent them. 

Lastly, for each bug, we identify its \textit{symptom} (such as crash, failure, slow performance, wrong output) and create a short label that summarises its \textit{root cause} (for example, "suboptimal kernel building"). During the labelling procedure, we kept the list of already generated labels available to enable their reuse when necessary. We have labelled the 529 bugs in 8 rounds (targeting on average 60 bugs per round) and had a conflict resolution meeting after each round. We carefully tracked all conflicts (108 in total) and how they were resolved.


\subsection*{Step 4. Expert Interviews}
There are limited public and open-source resources available for Hybrid Quantum-Classical architecture at the moment. While many frameworks are open-source, the data available about them and the instances of programs using them are scarce. We therefore decided to conduct expert interviews to enrich and validate our dataset. This step is organised into the following sub-steps: selecting the participants (4.1); designing the interviews (4.2); and conducting and processing the interviews (4.3). The result of the latter step is fed into a further survey (Step 5) and is incorporated into the final taxonomy (Step 6). 

\paragraph*{4.1. Selecting 
Interviewees} 
To ensure diversity in our panel of experts, especially between industry and academia, we selected and contacted experts working on Hybrid Quantum-Classical architectures at several leading quantum computing companies such as Quantum IBM, Quasar USA Quantum Blockchain Technologies, Google Quantum, and ClassicQ. We approached them through e-mail, as well as LinkedIn and ResearchGate’s private messaging systems. We received three positive responses and managed to secure another industry interview (a total of four) through personal contacts. One of the interviewees from the industry was identified by us due to their active participation in fixing several of the GitHub issues in our dataset, and we contacted them on the LinkedIn platform.The remaining selected experts (seven in total) are active in academia. Two of them have a strong connection with the industry. Academic profiles vary from PhD student to Professor, both from Computer Science and Physics Departments, with significant experience in Quantum Computing and especially in Hybrid Quantum-Classical architectures.To ensure a structured analysis, all experts were explicitly categorised based on their academic vs. industrial affiliation and disciplinary background. We directly asked participants about their expertise, current role, and experience during the interviews, as can be verified in the transcripts provided in our replication package (see the “Interview” folder). This categorisation allowed us to contextualise their feedback appropriately and ensured that insights from both academic and industrial perspectives were incorporated into the taxonomy refinement process.

\paragraph*{4.2. Designing the Interview Structure}

We opted for a semi-structured interview to have a flexible and adaptive structure to elicit as much information from the experts as possible \cite{seaman_qualitative_1999}. After establishing the scope of the study, and defining the key terms as advised in the pilot interviews, we opened with a broad question, namely, \textit{'What kind of problems have you experienced developing Hybrid Quantum-Classical architectures?'}. We tried to adapt the questions to the interviewees and re-direct the discussion to lead them to share the faults they have experienced. We prepared a script, available in the replication package, to ensure consistency in the content of each interview while being adaptive to elicit most information where the interviewee had more experience. When the interviewee mentioned any fault during the interview, we asked additional questions to identify all the characteristics required for our collection (in the library or not, symptoms, etc.)

\paragraph*{4.3. Processing the Interviews}
We conducted 11 interviews with interviewees from different academic and industrial profiles. To validate the structure of the interview, we conducted them in two rounds: a pilot round and a final round. 
The first two pilot interviews were held by two of the authors. We asked for the feedback of the interviewees to improve the interview process. The two comments received were to better define the context of the study, start the interview with a formal definition of Hybrid Quantum-Classical architectures and faults, and send the questions in advance. Both the interviews and the feedback discussions were transcribed and are available in the replication package. Once we ensured the final interview structure, the remaining 9 interviews were held by the first author only. Each interview was transcribed once the consent from the interviewee has been obtained. Two of the authors analysed the transcription and extracted the faults mentioned in it independently, to later cross-validate the results through a consensus meeting. The highlighted new faults or categories were extracted to be incorporated into the dataset. In these cases, all relevant passages of the transcripts were highlighted, and each fault extracted is associated with an exact timestamp in the transcription. 

\subsection*{Step 5. Final Validations}

\paragraph*{5.1. Search Validation}
We explored other quantum bug benchmarks \cite{zhao_bugs4q_2021}\cite{zhao_empirical_2023} and selected all bugs that satisfied our inclusion criteria. Given that those benchmarks include fixed bugs, they all comply with criteria 2 to 4, which means we selected all bugs that occurred in Hybrid Quantum Classical architectures. We verified that all these bugs were included also in our dataset. Bugs4Q \cite{zhao_bugs4q_2021} included only 3 bugs with a noise simulation, and none were Hybrid Quantum-Classical architecture faults. The second study \cite{zhao_empirical_2023} focused on Quantum Machine Learning bugs. To validate our dataset, we checked all benchmark faults complying with our four criteria in the benchmark repositories and ensured our dataset included them. We selected 71 bugs in their repository, all of which were indeed already captured in our results.  \\

\paragraph*{5.2. Validation Surveys}

We created a validation survey to ensure the relevance of our taxonomy. The expert participants for validation were recruited through our personal contacts and project partners.  This survey was sent to an independent group of 25 researchers and industrial experts who did not participate in our interviews, of which 7 answered. After gathering their background, the participants were asked if they had encountered faults/problems in each of the categories of our final taxonomy. A final open question checked whether the participant had encountered faults that did not fall into our categories to validate the soundness of our taxonomy. The survey's results can be found in the next section, and more details are available in the replication package.

\subsection*{Step 6. Final Taxonomy Construction}
Once the final dataset was compiled, two of the authors analysed the categories and sub-categories separately and made the labels consistent (in terminology and style). This stage aimed to harmonise the results and find common patterns in the labels that could be gathered as a category or a sub-category. Our starting point for the main categories was the general architectural components identified in Step 3 (3.2), but we updated them by considering the outputs of the interviews. 
As a result, a category was removed and another was added at this stage. The resulting taxonomy is depicted in Fig. \ref{FinalTax} and are further elaborated in the replication package. We reflect upon this result in the remainder of this paper, as well. 

\section{Results}
\label{results}
In this section, we provide an overview of the various results obtained in the process of gathering, structuring, and validating our dataset of real faults, including the dataset itself, the taxonomy, and the reflections based on the dataset, interviews and surveys leading to answers to our research questions.

\subsection{Faults Dataset}
Our datasets consist of 133 real faults from GitHub issues and 52 from the interviews. The interviews resulted in adding a full category of ``Conceptualisation'' that was not captured by the GitHub mining. After the interviews, we decided to merge the ``Ansatz Creation'' category as a sub-category of Quantum Circuits, because they pertain to the same components of the architecture. Moreover, we gathered all API-related faults into a new category. As a result, we finally ended up with seven top-level categories: Parametrisation, Conceptualisation, API, Optimisation, Quantum Circuit, Measurement and GPU. Table \ref{tab:database} gives an overview of the partition of the dataset in our final categories. The numbers separated by a plus sign represent the faults from the GitHub mining and the interviews respectively. 35\% of GitHub faults produce a wrong output, 3\% performance issues, and the remaining cause different crashes. Interviewees mentioned, however, that over 75 \% of faults led to wrong outputs, 8\% causing performance issues, and the remaining producing different crashes. Further labelling (of symptoms and causes) and a description and link to each fault are available in the \href{https://doi.org/10.5281/zenodo.18172880}{replication package}.

\begin{table}[hbt!]
    \centering
        \caption{Overview of the final result of faults included in the dataset: ``Level'' is the fault location in the Hybrid architecture; ``Faults'' is the total number of faults, differentiating betw.\ GitHub- and interview-sourced faults   (GitHub+Interviews); ``In Lib.'' denotes the faults in the libraries and ``Not in Lib.'' are those in the code using the libraries (both differentiating Github- + Interview-sourced faults).  }
    \begin{tabular}{@{}c|c|c|c|c}   
    \hline
        \textbf{Level} & \textbf{Faults} & \textbf{In Lib.} & \textbf{Not in Lib.} & \textbf{Sub-Categories} \\
    \hline
      Parametrisation & 22+8  & 10+2 & 12+6 & 8 \\
      \hline
      Conceptual & 0+17 & 0+0 & 0+17 & 6\\
      \hline
      API & 30+6 & 5+4 & 25+2 & 3 \\
      \hline
      Optimisation & 34+11 & 24+3 & 10+8 & 12 \\
      \hline
      Quantum Circuit & 24+3 & 21+1 & 3+2 & 8 \\
      \hline
      Measurement & 15+7 & 13+5 & 2+2 & 6 \\
      \hline
      GPU & 8+0 & 8+0 & 0+0 & 4 \\
      \hline
      \textbf{TOTAL} & 133+52 & 81+15 & 52+37 & 47 \\
      \end{tabular}
    \label{tab:database}
\end{table}

\subsection{Final Taxonomy}
\begin{figure}[p]
    \centering
    \adjustbox{angle=90, max height=\dimexpr\textheight-3\baselineskip\relax, max width=\textwidth}{
        \includegraphics{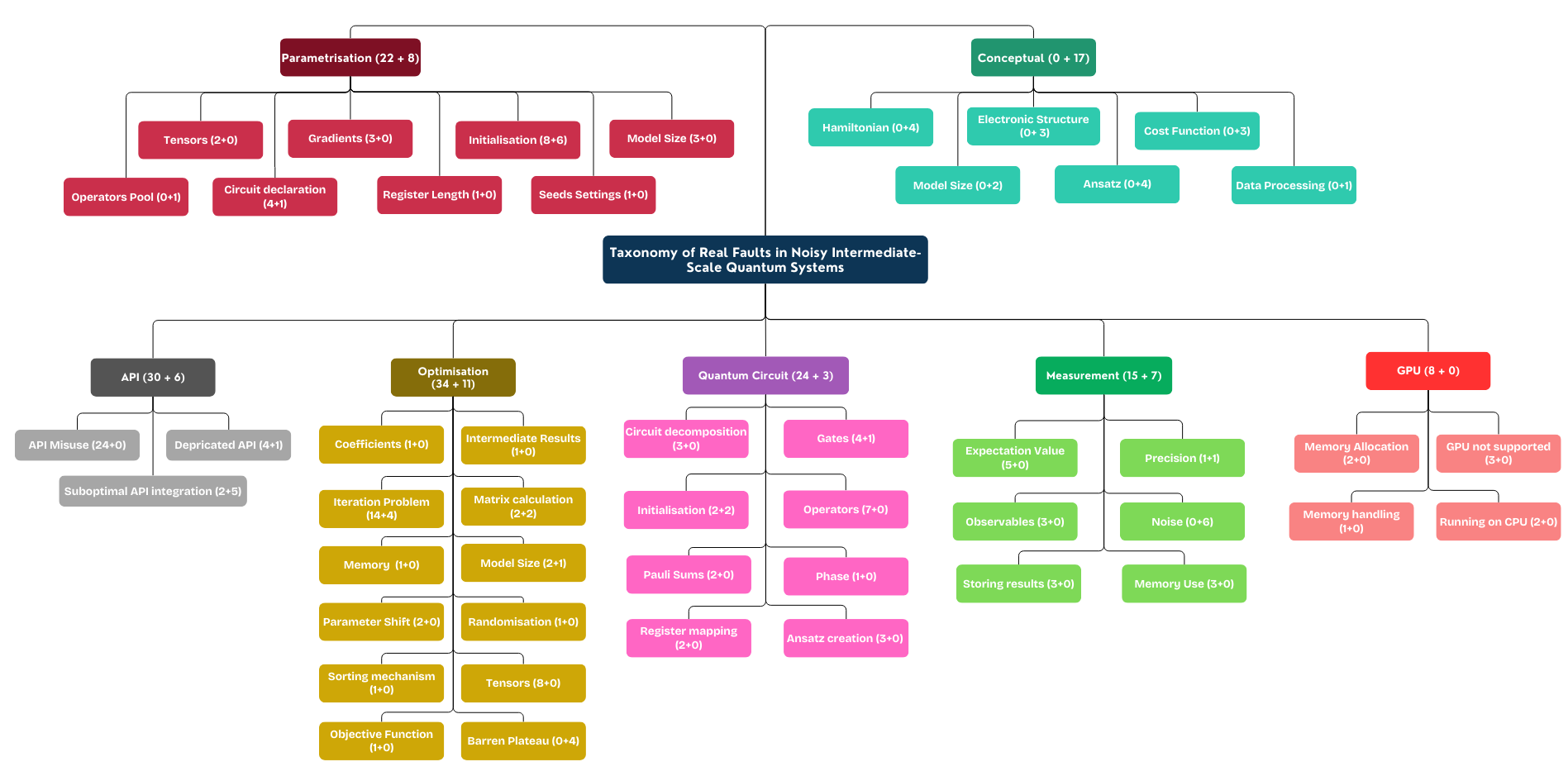}
    }
    \caption{The final taxonomy, organised into 7 top-level categories in darker colours, divided into 47 sub-categories in lighter shades of the same colour.}
    \label{FinalTax}
\end{figure}

The final taxonomy, displayed in Fig.~\ref{FinalTax}, is organised based on seven top-level categories, further divided into sub-categories. We decided to organise the taxonomy around categories representing the location of the fault, and sub-categories corresponding to the cause of this fault, to make sure it is relevant to both fault detection and fault localisation.
For each category and sub-category, we report the number of found real bugs; as before, the numbers are separated by a plus sign between the number of faults originating from the GitHub mining and those originating from the interviews,  respectively. As these numbers indicate, each subcategory contains multiple fault types. We have established 125 unique labels overall. However, due to space limitations, we could not demonstrate all of them in the taxonomy figure. In what follows next, we concisely present all categories and sub-categories. 


\paragraph{Parametrisation}
Parametrisation issues consist of all faults that occur at the initial phase of parametrisation of the quantum circuit and the optimiser, divided into 8 sub-categories listed in Table \ref{tab:parametrisation}. This category of faults was equally represented in GitHub and interviews, accounting for ca.\  15\% of faults in both sources. Faults in this category are usually caused by a wrong data type for parameters or a wrong initialisation and would result in both wrong outputs or crashes.

An example of a fault of this type is a \href{https://github.com/PennyLaneAI/pennylane/issues/900}{GitHub issue} in which a \textit{pennylane.numpy.tensor} type sequence is initialised as a parameter gate at the parametrisation phase, causing a Runtime Error. The key-words arguments are assumed to be non differentiable by default, which causes a crash with certain devices that  are not capable of supporting the application of a gate taking this type of argument. It was fixed by ensuring the arguments are marked as differentiable when passed to a QNode as a keyword argument. This fault has been placed in the \textit{Tensors} sub-category and was marked as a fault inside a library. 

\begin{table}[hbt!]
    \centering
        \caption{Faults happening at the parametrisation phase of the system, divided into 8 sub-categories. Typical cause is the most used label in this category, Nb is the number of faults from online resources and interviews respectively.}
    \label{tab:parametrisation}

    \begin{tabular}{@{}c|l|c}
       \textbf{Sub-Categories} & \textbf{Typical cause} & \textbf{Nb} \\
         \hline
       Tensors  & Wrong input type for tensors & 2+0 \\
         \hline
       Gradients  & Wrong initialisation type for the gradients & 3+0 \\
         \hline
        Parametric initialisation & Wrong initial point & 8+6 \\
          \hline
        Model Size & Suboptimal depth size for the model  & 3+0 \\
          \hline
        Operation Pool & Suboptimal number of operators in the pool & 0+1 \\
          \hline
        Circuit Declaration & Wrong qubit register declaration & 4+1 \\
          \hline
        Register Length & Wrong initialisation of the register & 1+0 \\
          \hline
        Seed Settings & Global seed is not set  & 1+0 \\
    \end{tabular}
\end{table}

\paragraph{Conceptual}
\label{conceptualisation}
This category emerged from the interviews and accounts for 33\% of the faults mentioned by experts as shown in Table \ref{tab:database}. It consists of the conceptual mistakes made while translating the domain knowledge into an instance of a Hybrid Quantum-Classical architecture. This translation process is far from trivial and several faults that may be introduced in the translation process are listed in Table 
\ref{tab:conceptual}. In Section \ref{vqa seq}, we have introduced the key concepts used in this category, such as Hamiltonians, which is the representation of a quantum system and corresponds to a specific electronic structure. During the iteration process, this electronic structure corresponding to the represented system can be broken, leading to erroneous behaviours. More details can be found in Tilly et al.'s detailed review of VQE \cite{tilly_variational_2022}.

For instance, four experts mentioned Conceptual faults introduced by suboptimal Ansatz design as discussed in Interview 5. A poor Ansatz design can significantly affect the optimisation process: it will lead to performance issues and wrong outputs. It can be the root cause of Barren plateaus issues. Yet, finding an optimal Ansatz design is very challenging even for domain experts, it is problem-specific, and still an active area of research \cite{tilly_variational_2022}.

\begin{table}[hbt!]
    \centering
    \caption{Faults happening at the conceptualisation phase, comprising 6 sub-categories. Typical cause is the most used label in this category, Nb is the number of faults from online resources and interviews respectively.}
    \label{tab:conceptual}
    \begin{tabular}{@{}c|l|c}
         \textbf{Sub-Categories} & \textbf{Typical cause} & \textbf{Nb} \\
         \hline
      Hamiltonian & Incorrect physical to qubit Hamiltonian transformation & 0+4\\
      \hline
      Electronic Structure & Broken electronic symmetries & 0+3 \\
      \hline
        Cost Function & Suboptimal cost function definition & 0+3 \\
        \hline
        Model Size & Suboptimal model size &  0+2 \\
        \hline
        Ansatz & Suboptimal Ansatz design & 0+4 \\
        \hline
       Data Processing  &  Mismatch between training data and parametric data &  0+1 \\
    \end{tabular}
\end{table}

\paragraph{API}
API issues pertain to either the usage of an API or an internal problem in the API itself. We found API misuse to be very common, representing over 20\% of the GitHub faults and 10\% of the interview faults as shown in Table \ref{tab:database}. This can be due both to the complexity of the frameworks and the lack of proper documentation for advanced features. The latter was mentioned several times during the interviews as a challenge in Hybrid Quantum-Classical architectures. Frameworks to implement customised Hybrid Quantum-Classical architectures are rare. Developers often need to integrate machine learning frameworks such as Tensorflow or PyTorch into quantum simulators such as Qiskit or Cirq. Faults in such an integration introduce crashes, failures, wrong behaviours, and performance issues. This category was initially a sub-category in several categories and was later consolidated into a coherent category, divided into 3 sub-categories listed in Table \ref{tab:api}. 

A common example of API issues is the use of Deprecated API as in the following \href{https://github.com/Qiskit/qiskit-aer/issues/1595}{example}. The simulators commonly used are constantly being refactored, and the documentation available online do not always lead to the latest version available. As a result, programs have a very short life expectancy and developpers constently need to adapt their code to the latest versions.

\begin{table}[hbt!]
    \centering
        \caption{Faults happening at the API level of the system, divided into 3 sub-categories. Typical cause is the most used label in this category, Nb is the number of faults from online resources and interviews respectively.}
    \label{tab:api}
    \begin{tabular}{@{}c|l|c}

         \textbf{Sub-Categories} & \textbf{Typical cause} & \textbf{Nb} \\
         \hline
      API Misuse & Developer Error & 24+0\\
      \hline
      Deprecated API & Calling old function & 4+1 \\
      \hline
        Suboptimal API integration  & Integration between frameworks & 2+5 \\
    \end{tabular}
\end{table}

\paragraph{Optimisation}
This category concerns the faults occurring in the classical code for optimisation. We initially named this category \textit{Training}, refering to Humbatova et al.'s Taxonomy of Real Faults in Deep Learning Systems \cite{humbatova_taxonomy_2019}. However, we realised at a later stage that \textit{Optimisation} defines better this category in the context of Hybrid Quantum-Classical architectures. This category is still labelled as \textit{Training} in the \href{https://doi.org/10.5281/zenodo.18172880}{replication package} to acknowledge the evolution of our study.
It is the biggest category of our taxonomy, both in the numbers of faults and of sub-categories, accounting for about 25\% of GitHub faults and 20\% of interviews` faults, divided into 12 sub-categories, listed in Table \ref{tab:training}. 
The large number and diversity of faults in this category are motivated by the corresponding complexity and size of the code: due to the noise in the current architectures, the amount of classical code for optimisation dominates the complexity and size of the quantum circuit. At each iteration, the measurements of the quantum circuit are stored as intermediate results, the cost function is updated, and the coefficients of each parametric gate and the weights of the Ansatz's layers are updated for the next iterations. A more detailed explanation is available in Tilly et al.'s review of VQE \cite{tilly_variational_2022}. 

An example of an optimisation fault from the \textit{Tensors} sub-category is a 
\href{https://github.com/tensorflow/quantum/issues/321}{GitHub issue} in which a fault occurs during the optimisation process where the data passed between each epoch do not have the same batch size. The fault was diagnosed by slowly building up the model and ensuring the shape inputs behave properly. It was fixed by setting a batch size appropriate to the training data. 

\begin{table}[hbt!]
    \centering
        \caption{Faults happening at the optimisation phase of the system, divided into 12 sub-categories. Typical cause is the most used label in this category, Nb is the number of faults from online resources and interviews respectively.}
    \label{tab:training}
    \begin{tabular}{p{3cm}|p{8cm}|c}
       \textbf{Sub-Categories} & \textbf{Typical cause} & \textbf{Nb} \\
         \hline
       Coefficients  & Wrong handling of unshifted coefficients & 1+0 \\
         \hline
      Handling of intermediate results  & Wrong calculation of intermediate operators &  1+0 \\
         \hline
        Iteration problems & Wrong handling of weights & 14+4 \\
          \hline
        Matrix Calculation & Mismatching ordering method for results  matrices  & 2+2 \\
          \hline
        Memory & Suboptimal memory consumption calculation & 1+0 \\
          \hline
        Model Size & Inefficient memory handling & 2+1 \\
          \hline
        Parameter Shift & Wrong handling of parameterized gates & 2+0 \\
          \hline
        Randomisation & Suboptimal random generation mechanism & 1+0 \\
         \hline
        Sorting Mechanism & Mismatch between sorting mechanism and associated probabilities &  1+0 \\
         \hline
        Tensors & Wrong input type passed to the tensors  & 8+0 \\
         \hline
        Objective Functions & Suboptimal objective function interaction with tensors &  1+0 \\
         \hline
        Barren Plateaus & Suboptimal Ansatz structure / Noisy Measurements &  0+4 \\
    \end{tabular}
\end{table}

\paragraph{Quantum Circuit}
In this category, elaborated in Table \ref{tab:quantum}, we gather the faults in the quantum circuit, divided into 8 sub-categories. These bugs are specific to the design of a quantum circuit and are caused by the incorrect design of the Ansatz, its parametric nature, or the iteration over the circuit. 

For instance, this \href{https://github.com/tensorflow/quantum/issues/176}{fault}, categorised as \textit{Ansatz Creation}, was introduced by the developer who did not place a Hadamard Gates Wall in the Ansatz before the model circuit, which broke the link between the input tensor and to the sample layer and led to incorrect outputs. The Ansatz was conceptualised properly, but wrongly implemented. This issue was found by debugging the outputs behaviour when modifying the Ansatz, until the origin of the wrong behaviour was found. 

\begin{table}[hbt!]
    \centering
       \caption{Faults originating from the quantum circuit, divided into 8 sub-categories. Typical cause is the most used label in this category, and Nb is the number of faults from online resources and interviews, respectively.}
    \label{tab:quantum}
    \begin{tabular}{@{}c|l|c}
       \textbf{ Sub-Categories} & \textbf{Typical cause} & \textbf{Nb} \\
         \hline
       Circuit decomposition  & Suboptimal handling of redundant qubits &  3+0 \\
         \hline
       Gates  & Wrong handling of custom gates  & 4+1 \\
         \hline
        Initialisation & Suboptimal kernel building &  2+2 \\
          \hline
        Operators & Suboptimal qubit reduction &  7+0 \\
          \hline
        Pauli Sums & Suboptimal handeling of PauliSum Operator & 0+1 \\
          \hline
        Phase & Wrong handling of global phases &  1+0 \\
          \hline
        Register Mapping & Incorrect handling of unusual types by register mapping &  2+0 \\
          \hline
        Ansatz Creation & Wrong input type passed to the layers &  3+0 \\
    \end{tabular}
\end{table}

\paragraph{Measurement}
In this category, listed in Table \ref{tab:measurements}, the issues related to the measurement phase are classified. They represent about 11\% of both GitHub and interview faults. Measurement is a crucial phase since it represents the transition from quantum information into classical information and many faults occur due to inappropriate choice of measurement (e.g., in grouping the observables) or transformation and storage of classical results (e.g., floating point operations or storing the variables). 

In this \href{https://github.com/tensorflow/quantum/issues/235}{issue}, a measurement imprecision accumulates with the circuit's size, leading to a wrong output. This is due to a sub-optimal extraction of the measurements introduced by the qsim version used when iterating over the qubits; hence it was labelled as an In Library fault. It was fixed by updating the iteration process ensuring the C++ version of qsim used and the referenced version do not mismatch.

\begin{table}[hbt!]
    \centering
        \caption{Faults originating from the measurements, divided into 6 sub-categories. Typical cause is the most used label in this category, Nb is the number of faults from online resources and interviews respectively.}
    \label{tab:measurements}
    \begin{tabular}{@{}c|l|c}
       \textbf{ Sub-Categories} & \textbf{Typical cause} & \textbf{Nb} \\
         \hline
       Expectation Value  & Suboptimal observables grouping strategy &  5+0 \\
         \hline
       Precision  & Wrong handling of floating points & 1+1 \\
         \hline
        Observables & Suboptimal measurement process &  3+0 \\
          \hline
        Noise & High noise level &  0+6 \\
          \hline
        Storing results & Suboptimal type for storing results &  3+0 \\
          \hline
        Memory use & Memory handling  & 3+0 \\
    \end{tabular}
\end{table}

\paragraph{GPU}
The issues related to the use of a GPU in the architecture are listed in Table \ref{tab:gpu}. 

This \href{https://github.com/Qiskit/qiskit-aer/issues/1647}{Segmentation Error example}, categorised as \textit{Memory Allocation}, is due to a Unitary matrix larger than the matrix buffer allocated to GPU. We expect this category to become more significant in the future since much work aims to further integrate GPUs into Hybrid Quantum-Classical architecture and quantum simulators.

\begin{table}[hbt!]
    \centering
        \caption{Faults originating from the GPU integration, divided into 4 sub-categories. Typical cause is the most used label in this category, Nb is the number of faults from online resources and interviews respectively.}
    \label{tab:gpu}
    \begin{tabular}{@{}c|l|c}
       \textbf{ Sub-Categories} & \textbf{Typical cause} & \textbf{Nb} \\
         \hline
      Memory Allocation  & Wrong memory allocation for GPU when the model is big &  2+0 \\
         \hline
       GPU not supported & Suboptimal GPU integration  &  3+0 \\
         \hline
        Memory handling & Memory Leak & 1+0 \\
          \hline
        Running on CPU & Implementation using CPU instead of GPU  & 2+0 \\
    \end{tabular}
\end{table}

\subsection{Validation Results}
Table \ref{tab:survey} displays the results of the validation survey. One of the seven participants answered "No" to the opening question 'Have you experienced Hybrid Quantum-Classical architecture faults', so we decided to exclude their contribution as it did not fit the required expertise - they only had some high-level theoretical knowledge of Hybrid Quantum-Classical architectures. Not all participants filled every sub-question about the severity and effort to solve per category; the answers received are summarised in Table \ref{tab:survey}. 4 participants declared having a Computer Science background, 2 came from Physics education, and one from Mathematics.

The results of the survey generally support the findings of the Github mining and the interviews: the identified categories of faults were experienced by at least 50\% of the participants, with 4 categories experienced by 83\% of them. All but one participant experienced at least half of the categories of faults. The aforementioned remaining participant has experienced 3 out of 7. When asked if they have encountered faults not present in the survey, two mentioned noise-induced errors - present as a sub-category of Measurements, and one mentioned a specific API problem that would fall into our API Misuse sub-category. The rest of the participants answered negatively.  
The survey provided additional insights that can be further used when developing testing, debugging, and repair techniques for these faults: critical faults seem to be concentrated in API, GPU, and measurement components. As we report below, our manual analysis also corroborates that the interfaces of components, particularly across the quantum and classical boundary are most prone to severe faults. The survey indicates that API faults are not only critical but difficult to resolve, while conceptualisation faults, which are mostly minor issues, are also very difficult to resolve due to the deep insight needed both in the domain and in the translation to quantum concepts. 
We acknowledge that the relatively small sample size of the survey limits its statistical power and generalisability. Recruiting experts in the highly specialised field of Hybrid Quantum-Classical computing remains a significant challenge, especially as survey participants were independent from the interviewees. Therefore, the survey is intended primarily as a qualitative validation to confirm consistency with findings from empirical analysis and expert interviews, rather than to support formal statistical conclusions. The following section details further our findings.

\begin{table}[hbt!]
    \centering
        \caption{Validation Survey Results. In each category, the ``Answer'' column indicates if the experts have encountered any such faults before. For those who have identified the faults in the category a further indication of the ``Severity'' (minor, major, or critical) of and the ``Effort to Resolve'' (low, medium, or high) the encountered fault are provided in the corresponding columns. }
    \label{tab:survey}
    \begin{tabular}{@{}c|@{}cc@{}|@{}ccc@{}|@{}ccc}
    \textbf{Categories}     & \multicolumn{2}{c|}{\textbf{Answers}}  & \multicolumn{3}{c|}{\textbf{Severity}} & \multicolumn{3}{c}{\textbf{Effort to Solve}} \\
                    & \textit{Yes} & \textit{No} & \textit{Min.} & \textit{Maj.} & \textit{Crit.} & \textit{Low} &  \textit{Med} & \textit{High} \\
    \hline
    Parametris.    & 3  & 3 & 1 & 0 & 0 & 1 &0 &  0\\
    \hline
    Conceptual     & 5 &  1 & 3 & 0 & 0 & 0 &2&  1\\
    \hline
    Measurement     & 4 & 2 &  2 & 0 & 3 & 1 & 2 & 0 \\
    \hline
    Optimisation     & 5 & 1 & 2 & 0 & 1 & 1 & 2 & 1\\
    \hline
    Quantum Circ.     & 3 & 3 & 1 & 0 & 1 & 1 & 0 & 1\\
    \hline
    API     & 5 & 1 & 1 &  0 & 3 & 0 & 1 & 2 \\
    \hline
    GPU     & 5 & 1 & 1 & 0 & 2 & 0 & 2 &  0 \\
    \end{tabular}
\end{table}

\subsection{Research Questions}
\paragraph{RQ1. Typical failure causes in Hybrid Quantum-Classical architectures}

Although Hybrid architectures are prone to regular quantum and classical faults, some faults are specific to this architecture. The iterative nature of such algorithms introduces further domain-specific challenges.

For example, imprecision in measurements accumulates over time and may cause wrong outputs that are difficult to detect. We observed that in the currently available frameworks for Hybrid Quantum-Classical architectures,  several type errors can be inadvertently introduced; these include simple type mismatches and rounding imprecision. Such type conversion and mismatch errors are significant in the interfaces of the different components in the architecture. They particularly prone to happen in the optimisation process while handling measurements, and passing intermediate results between Tensors.

The iterative nature of the architecture is resistant to noise to some extent and noise can even be exploited in such architecture to provide diversity in sampling. Noise-induced barren plateaus are, however, a common fault representing about 25\% of conceptual faults. Ansatz's structure affects the optimisation in different ways: poor Ansatz design can lead to performance issues as well as wrong outputs.

There are similar patterns of faults, mostly in parameterisation and optimisation, 
between ML and Hybrid Quantum-Classical architectures. However,  the introduction of quantum noise gives the classical optimisation process in Hybrid architecture (and its interface with the quantum circuit)  a unique nature, constituting 50\% of the faults mentioned in experts' interviews.

\begin{quote}
\textbf{\emph{Answer to RQ1:}} 
Over 40\% of faults in Hybrid Quantum-Classical architecture happen at the Parametrisation and Optimisation phase.  
Over 50\% of the faults mentioned in the expert interviews concern the classical optimisation component and its interface with the quantum circuit. 
A significant part of the novel faults contributed by the expert interviews concerns the conceptualisation of the problem from the domain knowledge into the Hybrid architecture. 

The results of the survey and our manual analysis of the real fault dataset both indicate that the interfaces between structural components of the architecture are the most fault-prone. We found noticeably input types mismatches, and version integration between the different components of the algorithms to be problematic.
\end{quote}

\subsection{RQ2. Contribution to the Taxonomy}

GitHub faults and faults mentioned in interviews complement each other nicely. Although GitHub issues typically gave insight into implementation issues, from the library itself or developer issues, interviews focused on a higher level of abstraction and comprehension errors. 

\begin{quote}
\textbf{\emph{Answer to RQ2:}}  
Our taxonomy was developed iteratively through a combination of empirical mining and expert validation. The initial structure emerged from a detailed analysis of GitHub issues, where faults were manually labeled based on their symptoms and technical origin. Expert interviews then played a crucial role in refining the categories, reorganising them to better reflect the practical understanding of fault causes.

For instance, an issue that would initially have been labelled as a ``Structural'' fault happening at the Ansatz level was reclassified during interviews as a \emph{conceptual error in Ansatz design}, drawing attention to deeper modelling challenges that are specific to Hybrid Quantum-Classical architectures. This shift highlighted the importance of differentiating between surface-level symptoms and root causes---an insight made possible by expert feedback. Moreover, the interviews revealed disciplinary nuances, such as computer scientists struggling more on conceptual faults and physicists on optimisation challenges, which helped balance the taxonomy to better capture the interdisciplinary nature of the field.

Importantly, the interviews also helped counterbalance the overrepresentation of computer science perspectives typically observed in online resources such as GitHub. Several interviewees, particularly from physics and mathematics backgrounds, explicitly mentioned not publishing or publicly reporting encountered faults---a bias that would otherwise limit the generalisability of the dataset. These qualitative insights ensured a more comprehensive and representative taxonomy.

The final taxonomy structure organises faults by their location (e.g., classical optimisation, quantum circuit, API layer) and further subcategorises them by cause (e.g., memory issue, circuit decomposition, suboptimal API integration). This approach ensures the taxonomy is both empirically grounded and practically relevant, supporting fault detection, localisation, and future adaptability to evolving Hybrid architectures.

\end{quote}

\section{Discussion}
\label{discussion}
In this section, we discuss some additional meta-observations from the interviews and surveys that did not fit in our taxonomy structure. 

\subsection{Hybrid Quantum-Classical faults vs. Faults in Deep Learning Systems}
Hybrid Quantum-Classical systems bear some similarities to deep learning systems \cite{rieser_tensor_2023}\cite{beer_training_2020}; thus, it  is  natural to compare our taxonomy with similar work in machine learning  \cite{humbatova_taxonomy_2019}. While several categories may seem similar at first, noticeably Optimisation, API, and GPU, the parallel is superficial: the root causes in those categories remain different in Hybrid Quantum-Classical architectures. Other categories are unique to the Hybrid Quantum-Classical architectures.  More specifically, for quantum machine learning (QML), although the optimisation process of the parametric gates is similar to training in deep learning, the encoding of the dataset into qubits is specific to QML and thus, the faults introduced in this architecture cannot be directly related to faults in deep learning.

\subsection{Computer Science vs.  Physics Expertise}
Quantum computing is inherently a multi-disciplinary field. While computer scientists and physicists often use the same Hybrid Quantum-Classical architectures, their approaches and the types of problems they encounter can differ significantly. This difference is clearly reflected in our interviews. Computer scientists tend to struggle more with conceptualisation-related faults, such as designing the Ansatz or effectively mapping a problem to a Hamiltonian, where the physical properties and noise characteristics of quantum hardware play a critical role. Physicists, on the other hand, often face more challenges during the optimisation phase, which requires abstract reasoning about algorithms and software design.

These disciplinary differences also influence how faults are documented and shared within the community. Our interviews revealed that computer scientists are generally more likely to publish faults and open issues on public platforms like GitHub, which is reflected in the dominance of computer science perspectives in data mined from such sources. Conversely, experts from physics, chemistry, and mathematics backgrounds often do not publish faults or open issues publicly, either due to the sensitive nature of experimental setups or different cultural norms in their fields. This discrepancy was crucial to uncover through expert interviews, as it highlighted the limitations of relying solely on public datasets for taxonomy development and motivated our inclusion of a conceptualisation category based on the unique insights from these interviews. Overall, this underlines the importance of combining empirical mining with expert consultation to capture a fuller, more balanced picture of the fault landscape in Hybrid Quantum-Classical architectures.

\subsection{Manual Debugging Methods}
Two approaches were mentioned to investigate and resolve the identified faults: one focusing on the optimisation process, and another on the problem conceptualisation. In the first method, the developer detects non-convergence, and first focuses on debugging the implementation and behaviour of the gradients. They would manually debug each gradient at every iteration to check if the results are correctly passed and to detect where the error may originate from. The second approach is more problem-specific and was mentioned by several experts with a Physics background. In case of a wrong behaviour of the system, they would first focus on their problem definition and manually ensure that the Hamiltonian properly represents the problem. The next step would be to manually ensure that the Hamiltonian is correctly mapped into qubits. Mapping a Hamiltonian into qubits is not trivial, and finding an efficient mapping is an open question  \cite{tilly_variational_2022}. They would then focus on the structure of the Ansatz, and ensure no electronic symmetry is broken (see  Section \ref{conceptualisation}). Only then, if no bugs are found in the previous steps, they would proceed to similar steps as the first approach. In Interview 5, the expert gives a detailed explanation of such a procedure. Many of these steps could be automated in the future.

\subsection{Common Problems}
The semi-structured interviews led us to discuss experts' problems beyond our taxonomy. A topic mentioned frequently was the lack of accurate and up-to-date documentation. Quantum platforms tend to mostly document their basic functionalities, leaving advanced features undocumented. Moreover, experts miss a flexible Hybrid Quantum-Classical architecture to experiment on. 
Tensorflow Quantum \cite{broughton_tensorflow_2021} was introduced with this purpose, but stopped being maintained. Several interviewees mentioned having to implement their connections between machine learning frameworks and quantum simulators, which often introduce compatibility errors and performance issues. With the growing interest in Hybrid architecture, researchers and developers would benefit from a dedicated platform that would strongly integrate all components of such architectures and ensure robust and reliable implementations, particularly where quantum and classical parts interact. Although common platforms such as Qiskit and Pennylane allow some support for common NISQ algorithms, such as VQA, experts lack more flexible architecture and templates for various components - typically Ansatz and Optimisation, with clear guidelines for the potential applications and customisations.

\subsection{Practical Applications and Guidelines for Use}
\label{sec:usage-guidelines}

Hybrid Quantum-Classical architectures are an emerging area of research with increasing practical significance. However, prior studies have yet to empirically capture the specific nature and limitations of faults in such architectures. Our work addresses this gap by introducing the first structured taxonomy of real-world faults in Hybrid Quantum-Classical architectures, grounded in actual implementations and validated through expert feedback. This constitutes a novel theoretical contribution to the field of quantum software engineering, offering a foundational understanding of fault landscapes in Hybrid architectures.

Beyond its theoretical value, our taxonomy is supported by a curated dataset of 133 empirically observed faults, enabling several practical applications:

\begin{itemize}
    \item \textbf{Grounding testing and debugging tools} in real-world hybrid faults, thus aligning tool development with the actual challenges faced by developers.
    
    \item \textbf{Enabling fault injection and simulation} using empirically derived fault patterns to benchmark the effectiveness of quality assurance techniques.
    
    \item \textbf{Supporting education and documentation} by illustrating recurring pitfalls in hybrid program implementation, thereby aiding onboarding and training.
    
    \item \textbf{Guiding future empirical studies} through a reusable and extensible classification framework, facilitating comparative analysis across platforms and toolchains.
\end{itemize}

We plan to integrate this taxonomy and dataset into a toolbox for generating faulty implementations of Hybrid architectures. This resource will support the empirical evaluation of testing, debugging, and automated repair techniques. We are conducting a comprehensive analysis of our findings, derived from both the dataset and the interviews, to develop testing oracles capable of identifying incorrect behaviour that do not result in crashes for the different fault types by classifying their symptoms. These oracles constitute a critical step toward advancing the understanding of hybrid faults and provide a foundation for future research in Quantum Software Testing, both at the theoretical and empirical levels. By making both the taxonomy and dataset publicly available, we aim to promote reproducibility and accelerate the advancement of robust, testable Hybrid Quantum-Classical architectures.

Furthermore, this field is inherently interdisciplinary, drawing on deep expertise from both quantum physics and software engineering. As such, our taxonomy and dataset can serve as a bridge between communities: helping researchers with physics backgrounds gain insight into practical software engineering challenges, and conversely, assisting software engineers in understanding the fundamental constraints and behaviours of quantum systems when embedded into classical computations.

The expert interviews conducted as part of this study also provide valuable qualitative data, offering diverse perspectives on how Hybrid Quantum-Classical architectures are developed and maintained in practice. These interviews can serve as a resource for researchers and practitioners seeking a deeper understanding of the ecosystem, tooling gaps, and fault patterns encountered by different categories of users.

\paragraph{Practical Guidelines}
To support both practitioners and researchers working with Hybrid Quantum–Classical architectures, we propose four main use cases for our taxonomy and dataset. These use cases illustrate how structured knowledge of causes (faults) and their effects (symptoms and failures) can drive improved system design, testing, and debugging practices.
\begin{enumerate}
\item \textbf{Categorising Faults in Hybrid Architectures and Analysing Cause (Fault) — Effect (Symptom/Failure) Relationships:}
The taxonomy organises common failure causes across the life cycle of hybrid architectures, from parametrisation and conceptualisation to API, optimisation, and circuit implementation. For example, at the \textit{parametrisation} stage, developers often introduce faults such as wrong input types for tensors or incorrect initial points in parametric initialisation; in the \textit{conceptualisation} stage, faults caused by suboptimal Ansatz design or broken electronic symmetries can arise. By linking these causes to their typical effects (symptoms and failures), researchers can study relationships between specific mistakes and the effects they trigger. This knowledge can inform the design of better development practices, guide education efforts, and support future cause-prediction or fault-prediction models.
\item \textbf{Developing Specialised Oracles to Trigger Effects Caused by Faults:}  
The dataset supports the design of test oracles aimed at detecting causes that lead to silent but incorrect behaviours — that is, faults that do not cause program crashes. For instance, a fault in Ansatz design can produce a \textit{barren plateau} that silently degrades optimisation performance, while high noise in measurement processes can cause biased outcomes without explicit error messages, and violations of physical constraints, such as particle number or spin, may go unnoticed. By implementing oracles that monitor gradient magnitudes, symmetry preservation, and statistical properties of outputs, and integrating them into continuous testing pipelines, silent failures can be detected early. These oracles improve reliability and provide actionable feedback for hybrid developers.

\item \textbf{Test Case Generation Techniques:}  
Insights from the taxonomy guide the development of input generation methods and test suite optimisation. Strategies include:  
\begin{itemize}
    \item \emph{Parametric Variation:} systematically vary initial tensor parameters, circuit depths, or optimizer hyperparameters to expose faults in parametrisation and optimisation phases.  
    \item \emph{Circuit Structure Manipulation:} explore alternative Ansatz, qubit mappings, or distributed circuit decompositions to uncover implementation and parallelisation faults, including inconsistencies when recombining subcircuits.  
    \item \emph{Constraint Violation Probes:} generate states that intentionally challenge physical constraints (e.g., particle number, spin, symmetry) to test the sensitivity and effectiveness of physics-consistency oracles.  
    \item \emph{Noise Stress Testing:} inject controlled stochastic gate errors or simulate decoherence to quantify robustness and identify brittle components.  
    \item \emph{Hybrid Workflow Coverage:} design sequences that combine classical preprocessing, quantum execution, and classical postprocessing in varied configurations to expose integration or optimizer-induced failures.  
\end{itemize}  
Linking test generation strategies to cause categories enables practitioners to focus on the most fault-prone aspects of hybrid architectures and produce efficient, targeted, and reproducible test suites that go beyond naïve random sampling.

\item \textbf{Automatic Fault Localisation and Repair:}  
The structured mapping from faults to effects supports automated debugging tools. Observed failure patterns can be traced back to likely root causes. For example, a fault in parameter-shift rule implementation during optimisation often manifests as low-fidelity outputs, while memory-handling causes — such as improper GPU allocation — may cause performance degradation.  Combined with specialised oracles and targeted test cases, this information supports semi-automatic or fully automated repair strategies, such as suggesting alternative Ansatz, modifying circuit decomposition strategies, or adjusting optimizer settings. This approach can shorten debugging cycles, improve reproducibility, and enhance software reliability in large-scale hybrid architectures.
\end{enumerate}

\paragraph{Potential Application Areas}

The proposed taxonomy and dataset are broadly applicable across key domains that rely on Hybrid Quantum–Classical architectures. These include: (i) \textit{quantum chemistry}, where variational algorithms such as VQE are used to approximate molecular energies; (ii) \textit{drug discovery and repurposing}, where hybrid workflows optimise quantum–classical pipelines for molecular similarity and binding prediction; (iii) \textit{optimisation and operations research}, encompassing quantum approximate optimisation algorithms (QAOA) and hybrid solvers used for scheduling, routing, and resource allocation. In these contexts, systematic understanding of faults and their effects can improve robustness, reproducibility, and the interpretability of hybrid computational outcomes.

\section{Threats to validity}
\label{threats to validity}
\paragraph{Internal Threats} 
An internal threat to validity are the potential biases in data processing. To mitigate it, we cross-validated the results between two authors, as well as with two independent groups of domain experts through interviews and a validation survey, respectively. The interview questions remained general to avoid leading the interviewees towards our taxonomy. We insisted on balancing the profiles of interviewees to represent diverse views. We decided to represent our dataset in a structural taxonomy. However, different representations may be considered, for instance around the symptoms and common patterns in the labels, creating a multi-view taxonomy in the future. 

\paragraph{External Threats}
Hybrid Quantum-Classical architectures are relatively new, and their implementations undergo constant revisions. It is challenging to capture structural faults not solely caused by the evolving implementations. It is also challenging to know to what extent our taxonomy can be generalised to future architectures. While we expect that noise-related problems may be mitigated, other faults in this taxonomy inherent to the Hybrid Quantum-Classical architecture should remain relevant. Recent algorithmic paradigms such as the realisation of a quantum neural network using repeat-until-success circuits in a superconducting quantum processor~\cite{moreira2023realization}, exemplify the pace at which the field evolves. While these approaches introduce new design patterns, we believe that several fault types we documented—such as parameterization faults, quantum circuit design errors, and API misuse—are likely to remain relevant for such emerging models. Furthermore, similar fault types already appear in our dataset, particularly in contexts related to quantum machine learning. This suggests that certain structural weaknesses in Hybrid Quantum-Classical software development persist across algorithmic generations. Even though specific faults from these new paradigms were not directly observed during our study period, our taxonomy provides a principled framework that can help identify, classify, and reason about faults in these evolving algorithmic models.

\section{Conclusions}
\label{conclusion}

We performed a structured study of real faults reported in open-source Hybrid Quantum-Classical architectures. Current instances of such architectures are concentrated around NISQ algorithms such as  Variational Quantum Eigensolvers and
Quantum Approximation Optimization Algorithms. We designed and validated a search query and defined rigorous inclusion/exclusion criteria. We analysed
5000 closed issues on GitHub and selected 529 of them. 
 We validated the results externally in two rounds with two independent groups of experts through semi-structured interviews and surveys, respectively. These led to updates and additions to the documented faults and their classification.

We plan to incorporate our dataset into a toolbox for generating faulty implementations of Hybrid Quantum-Classical architectures. This will serve as a means for evaluating future techniques for testing, quality assurance, and repair of such architectures, which is a direction of our ongoing research. We would like to further investigate other views that can be used to classify our dataset of real faults. As future work, we aim to build on these findings—both from the fault dataset and the interviews—to develop testing oracles capable of detecting incorrect behaviours that do not result in crashes. Such oracles, directly informed by this taxonomy, will represent an important step toward more reliable and robust Hybrid Quantum–Classical architectures.
\break

\noindent \textbf{Replication Package}
\textit{All the data used to build our taxonomy is available in the following  \href{https://doi.org/10.5281/zenodo.18172880}{repository}.
}
\\

\noindent\textbf{Acknowledgments. } Avner Bensoussan and Mohammad Reza Mousavi  have been partially supported by the UKRI
Trustworthy Autonomous Systems Node in Verifiability, Grant
Award Reference EP/V026801/2, EPSRC project on Verified Simulation for Large Quantum Systems (VSL-Q), grant
reference EP/Y005244/1 and the EPSRC project on Robust
and Reliable Quantum Computing (RoaRQ), Investigation
009 Model-based monitoring and calibration of quantum
computations (ModeMCQ), grant reference EP/W032635/1. 
Gunel Jahangirova and Mohammad Reza Mousavi have been partially supported by the 
ITEA/InnovateUK projects GENIUS, grant reference 600642, and 
ITEA/InnovateUK GreenCode, grant reference 600643.

\printbibliography

\end{document}